\documentclass[a4paper,american,aps,citeautoscript,floatfix,longbibliography,pdftex,pre,
superscriptaddress,showpacs,preprint]{revtex4-1}

\pdfoutput=1
\usepackage{adjustbox}
\usepackage{natbib} 
\usepackage{hyperref}

\usepackage{amsmath,amssymb,amsmath}
\usepackage{graphicx}
\usepackage{dcolumn}
\usepackage{bm}
\usepackage{color}
\usepackage{ulem}

\makeatletter
\def\paragraph{\@startsection{paragraph}{4}{10pt}{-1.25ex plus -1ex minus -.1ex}{0ex plus 0ex}{\normalsize\textit}}
\renewcommand\@biblabel[1]{#1}
\renewcommand\@makefntext[1]%
{\noindent\makebox[0pt][r]{\@thefnmark\,}#1}
\DeclareRobustCommand\onlinecite{\@onlinecite}
\def\@onlinecite#1{\begingroup\let\@cite\NAT@citenum\citealp{#1}\endgroup}
\def\tagform@#1{\maketag@@@{\ignorespaces#1\unskip\@@italiccorr}}
\let\orgtheequation\theequation
\def\theequation{(\orgtheequation)}
\makeatother

\begin{document}

\title{Energy transfer mechanisms in a dipole chain: From energy equipartition to the formation of breathers}

\author{Alexandra Zampetaki}
\affiliation{Zentrum f\"ur Optische Quantentechnologien, Universit\"at
  Hamburg, Luruper Chaussee 149, 22761 Hamburg, Germany}

  \author{J. Pablo Salas}
\affiliation{\'Area de F\'{\i}sica, Universidad de La Rioja, 26006 Logro\~no, La Rioja, Spain}

\author{Peter Schmelcher}
\affiliation{Zentrum f\"ur Optische Quantentechnologien, Universit\"at
  Hamburg, Luruper Chaussee 149, 22761 Hamburg, Germany} 
\affiliation{The Hamburg Center for Ultrafast Imaging, Luruper Chaussee 149, 22761 Hamburg, Germany}

\date{\today}

\begin{abstract}
We study the energy transfer in a classical
dipole chain of $N$ interacting rigid rotating dipoles.
The underlying  high--dimensional potential energy landscape is analyzed in particular by
determining the equilibrium points and their stability in the common plane of rotation.
Starting from the minimal energy configuration, the response of the chain to excitation of a
single dipole is investigated. Using both the linearized and the exact
Hamiltonian of the dipole chain, we detect an approximate excitation energy threshold
between a weakly and a strongly nonlinear dynamics. 
In the weakly nonlinear regime, the chain
approaches in the course of time  the expected energy equipartition among the dipoles.
For excitations of higher energy, strongly localized excitations appear whose trajectories in time are either periodic or
irregular, relating  to the well-known discrete or chaotic breathers, respectively.
The phenomenon of spontaneous formation of domains of opposite polarization and phase locking is found 
to commonly accompany the time evolution of the chaotic breathers.
Finally, the sensitivity of the dipole chain dynamics to the initial conditions  is studied as a function of the initial 
excitation energy by computing a fast chaos indicator. The results
of this study confirm the aforementioned
approximate threshold value for the initial excitation
energy, below which the dynamics of the dipole chain is regular and above
which it is chaotic.
\end{abstract}
\maketitle

\section{Introduction}
\label{sec:introduction}

The first numerical study of the energy transport in a one-dimensional (1D) nonlinear oscillator chain, known
as the Fermi-Pasta-Ulam (FPU) model \cite{FPU,A780,lady} has been performed already in nineteen fifty five. The results of this numerical experiment were found to contradict the reasonable assumption that in the presence of a non-linear coupling between the oscillators the system would thermalize, i.e. an initial excitation of a single mode of the system would 
become equally distributed between all the modes of the chain. In particular, the numerical results
showed a persistent
recurrence of the energy to the initially excited mode, preventing the system from reaching equipartition up to long times.
It has been soon realized that the origin of such a behavior were the nonlinear interaction terms, a fact that established the study of the  energy exchange in
discrete nonlinear lattices of oscillators as an active field of research in few- and many-body dynamics, see for 
instance Refs.~\cite{Ford,A897,A780,A842,Gallavotti,Mussot,Bambusi,penati}.
Most attention has been paid to 1D oscillator chains with a cubic or quartic nonlinear coupling, the so-called FPU-$\alpha$ and FPU-$\beta$
models, respectively,  \cite{A780}. Already studies of the energy transfer
 in these simple FPU models have provided interesting results, such as the existence
of thresholds for stochasticity and therefore for equipartition \cite{A842,A899,A782}, as well as the
discovery of phenomena
of energy localization in discrete \cite{A866,A867,A868} or chaotic breathers \cite{A867,A851,A837}.

Beyond the theoretical FPU-like models, the mechanism of energy exchange 
is an important subject of investigation in  microscopic systems such as molecules, interacting via
Coulomb, dipole-dipole or van-der-Waals interactions. These fundamental interactions appear in
different research disciplines including physics, chemistry, biology and
material sciences, with applications covering such diverse topics as the photosynthesis of plants and 
bacteria~\cite{van,engel,fotosintesis,wu,Burghard}, 
the emission of light of  organic materials~\cite{Saikin,Melnikau,Qiao},
molecular crystals~\cite{Davydov,Silbey,Wright} or artificial molecular rotors~\cite{A829}.
Moreover the advances in current technology have 
allowed the trapping and confinement of cold molecules in
optical lattices where the positions of the molecules are fixed 
and their mutual interactions (e.g. dipole--dipole) usually masked by thermal fluctuation, become prominent~\cite{krems,weidemuller}.
Along these lines polar diatomic molecules trapped in optical lattices,  exhibit due to their strong dipole-dipole interaction a particularly
interesting quantum many-body behavior leading to novel
structures and collective dynamics~\cite{A753,rey,Lewenstein}.

Within the framework of classical mechanics, confined polar diatomic molecules can be considered as lattices of
rigid dipoles. Following this approach, Ratner and
co-workers~\cite{Ratner1,Ratner2,Ratner3}  have studied the energy transfer in
chains of interacting rotating rigid dipoles in various planar configurations. 
Already the simplest
two--dipole chain, recently revisited in~\cite{PRE2017}, was found to display a rich dynamical behavior with a complicated phase space. Increasing the number of dipoles the 
energy transfer was shown to yield the formation of solitons or the emergence of chaoticity
~\cite{A898}.

The objective of this paper is to provide further insights in the energy transfer mechanisms
of a 1D chain of rotating classical dipoles. In such a chain the dipoles
are assumed to be fixed in space, interacting through nearest neighbor (NN) interactions
and rotating in a common plane.  The interaction potential of even this simplified rigid-rotor model is
found to be quite complex, supporting various equilibrium points including a minimum, a maximum
and different saddle points. Considering the system in its ground state (GS) configuration (minimum)
with a single dipole excited initially possesing a certain  amount of
kinetic energy we study the transport of the excess energy. We find that for increasing excess
energy the degree of chaoticity of the energy transfer increases, passing through a weakly nonlinear
and a highly nonlinear regime. Although in the former regime after some time the energy is almost 
equally partitioned among the dipoles of the chain, for high enough excitation energies the  energy
diffusion is prohibited, giving place to different energy localization patterns dictated by a strong
nonlinearity. Among those patterns we can distinguish cases in which two domain walls,
separating domains of dipoles with different polarization, are formed spontaneously and
move irregularly in time. It turns out that the emergence of such patterns can be linked to the lowest energy saddle point of the interaction potential of the dipole chain. 
Moreover, for a large excitation energy, the dipole chain 
displays a strong sensitivity to the initial conditions, signifying its chaotic nature.
We quantify the chaoticity of the system for different values of the excitation energy using a
fast Lyapunov indicator.

The structure of the current paper is as follows. In Sec. II we present the Hamiltonian
and the equations  of motions of the dipole chain and discuss their equilibria.
Linearizing these equations of motion around the GS,
we arrive at the corresponding linear system whose properties are analyzed in Sec. III. 
The results for the energy transfer of a localized excitation are presented in Sec. IV and Sec. V. 
In particular, Sec. IV deals with the propagation of a low-energy excitation in 
the so-called weakly nonlinear regime, whereas Sec. V discusses the case of higher energy excitations in which 
the nonlinearity of the system is enhanced, leading to its chaotic behavior which is
quantified by the Orthogonal Fast Lyapunov Indicator.
Finally we provide our conclusions in Sec. VI.

\section{The Hamiltonian and the Equilibrium points}
We consider a linear chain of $N$ identical rigid dipoles of electric dipole moment 
${\bf d}_i=d \ {\bf u}_i$, which are fixed in space, separated by a constant distance $a_l$, and located
along the $X$-axis of the Laboratory Fixed Frame (LFF) $XYZ$.
The unit vectors ${\bf u}_i=(u_{xi}, u_{yi}, u_{zi})$ determine the orientation of each dipole
subjected to the holonomic constraint $\left|{{\bf d}_i}\right|^2=d_{xi}^2+d_{yi}^2+d_{zi}^2=d^2$.
The potential energy ${\cal V}_{ij}$ between each pair $(i,j)$ of rotors due to the
mutual dipole-dipole interaction (DDI) is given by \cite{A753}
\begin{equation}
\label{dipoleInteraction}
{\cal V}_{ij}=\frac{1}{4\pi\epsilon_0}
\frac{({\bf d}_i \cdot {\bf d}_j) \ r_{ij}^2 - 3 \ ({\bf d}_i \cdot  {\bf r}_{i,j}) \ ({\bf d}_j \cdot  {\bf r}_{i,j})}{r_{ij}^5}
\end{equation}
with $ {\bf r}_i=\left(x_i,y_i=0,z_i=0\right)$, ${\bf r}_{i,j}={\bf r}_i-{\bf r}_j$ and $r_{ij}=|{\bf r}_{i,j}|$.
 
Here we assume periodic boundary conditions (PBC) in the linear chain and we take into account
only interactions between nearest neighbors (NN), the total interaction potential ${\cal V}$
of the system reads
\begin{equation}
\label{TotalDipoleInteraction}
{\cal V}=\sum_{i=1}^N \ \frac{1}{4\pi\epsilon_0 a_l^5}
\bigg[({\bf d}_i \cdot {\bf d}_{i+1}) \ a_l^2 - 3 \ ({\bf d}_i \cdot  {\bf r}_{i,i+1}) \ ({\bf d}_{i+1} \cdot  {\bf r}_{i,i+1})\bigg].
\end{equation}
It is convenient to express the total interaction potential ${\cal V}$ in terms of
the Euler angles $(0\le \theta_i\le \pi, 0\le \phi_i<2\pi)$ of each rotor, such that  \ref{TotalDipoleInteraction}
takes the form
\begin{equation}
\label{dipoleInteraction2}
{\cal V}(\theta_i, \phi_i)=\alpha
\sum_{i=1}^N \bigg[\cos\theta_i\cos\theta_{i+1} +\sin\theta_i\sin\theta_{i+1} (\sin\phi_i\sin\phi_{i+1} -
2 \cos\phi_i\cos\phi_{i+1})\bigg],
\end{equation}
where $\alpha=d^2/4\pi\epsilon_0 a_l^3$ is the strength of the DDI. 
Note that the well-known stable {\sl head-tail} configurations of the dipoles appear
at $\theta_i=\pm\pi/2$ and $\phi_i=0, \pi$.
The rotational dynamics of the dipole chain is described by the Hamiltonian
\begin{equation}
\label{hamiEuler}
H =\sum_{i=1}^N \frac{1}{2 I}  \bigg[ p_{\theta_i}^2+
 \frac{p_{\phi_i}^2}{\sin^2\theta_i}\bigg]+{\cal V}(\theta_i, \phi_i),
\end{equation}
where $I$ is the moment of inertia of each dipole.
The Hamiltonian \ref{hamiEuler}
defines a dynamical system with $2N$ degrees of freedom $\{(\theta_i,p_{\theta_i}),(\phi_i,p_{\phi_i})\}_{i=1}^N$ where $p_{\theta_i},p_{\phi_i}$ 
denote the conjugate momenta of $\theta_i, \phi_i$ respectively.
From the corresponding Hamiltonian equations
of motion, it is easy to see that the
manifold ${\cal M}$ of codimension $N$ given by
\begin{equation}
\label{manifoldM}
{\cal M} = \lbrace(\theta_i, p_{\theta_i}) \ | \ \phi_i= 0, \pi \ \mbox{and} \ p_{\phi_i}=0\rbrace,
\end{equation}
is invariant under the dynamics. On this manifold, the number of degrees of freedom of the
system reduces to $N$ and the Hamiltonian \ref{hamiEuler} becomes
\begin{equation}
\label{eq:hamFPU}
H =\sum_{i=1}^N \frac{p_{\theta_i}^2}{2 I} +\alpha
\sum_{i=1}^N \bigg[\cos\theta_i\cos\theta_{i+1}-2 \sin\theta_i\sin\theta_{i+1}\bigg],
\end{equation}
and the rotational motion of the dipoles is restricted to a given common polar plane of
constant azimuthal inclination 
$\phi_i=0,\pi$ where the $N$ polar angles $\theta_i$ vary
 in the interval $[-\pi, \pi)$. From now on, we focus on the planar dynamics arising from 
the Hamiltonian \ref{eq:hamFPU}. It is worth noticing that the Hamiltonian~\ref{eq:hamFPU}
is structurally stable in the sense that, for weak enough perturbations away from
the manifold ${\cal M}$ and around the {\sl head-tail}
configuration, the dynamics takes place in the neighborhood of this
configuration, which is the absolute minimum
of the potential ${\cal V}(\theta_i, \phi_i)$.

As mentioned above the  stable {\sl head-tail} configurations of the dipoles in the manifold ${\cal M}$
appear for  $\theta_i=\pm\pi/2$. 
For the sake of simplicity, we choose to move these equilibrium configurations  to the origin
$\theta_i=0$ and to $\theta_i=\pi$ respectively. To this end, we introduce the following canonical transformation
between the previous $(\theta_i, p_{\theta_i})$ and the new  $(x_i, p_i)$ coordinates
\begin{equation}
\label{eq:cano}
x_i=\theta_i-\pi/2,\qquad p_i=p_{\theta_i}.
\end{equation}
Employing this transformation, the Hamiltonian \ref{eq:hamFPU} obtains the form
\begin{equation}
\label{eq:hamFPU1}
H =\sum_{i=1}^N \frac{p_i^2}{2 I} +\alpha
\sum_{i=1}^N \bigg[\sin x_i\sin x_{i+1}-2 \cos x_i\cos x_{i+1}\bigg],
\end{equation}
where $p_i=I \ dx_i/dt$.
Taking into account that our dipole chain model of Eq. \ref{eq:hamFPU1} amounts essentially to the rigid rotor model used for the study
of the dynamics of $N$ interacting
polar diatomic molecules \cite{Ratner1,Ratner2,Ratner3}, we find it convenient to express
the energy, i.e. the Hamiltonian \ref{eq:hamFPU1}, in units of the molecular rotational constant $B=\hbar^2/2 I$. 
To this end we 
define a new dimensionless time $t'=t/t_B$ with $t_B=\hbar/\sqrt{2} B$ whose use  leads us
to the following (dimensionless) Hamiltonian 
\begin{equation}
\label{eq:hamFPU2}
E' =\frac{H}{B} =\sum_{i=1}^N  \frac{p_i'^2}{2}+
\chi \sum_{i=1}^{N} \bigg[\sin x_i \sin x_{i+1}
-2 \cos x_i \cos x_{i+1} \bigg],
\end{equation}
\noindent
where $p_i'=dx_i/dt'$ and
$\chi=\alpha/B$ is a dimensionless parameter controlling the dipole interaction.
Besides the reduced energy $E'=H/B$, the dynamics of the system described by \ref{eq:hamFPU2}
depends also on the dipole parameter $\chi$.
However,  this dependence can be removed by further rescaling the time, introducing
$t'' = \sqrt{\chi} \ t'$. In terms of time $t''$, the
Hamiltonian~ \ref{eq:hamFPU2} reads
\begin{equation}
\label{eq:hamFPU3}
E \equiv {\cal H}=\frac{H}{B \chi} =\sum_{i=1}^N  \frac{p_i''^2}{2}+
\sum_{i=1}^{N} \bigg[\sin x_i \sin x_{i+1}
-2 \cos x_i \cos x_{i+1} \bigg] ,
\end{equation}
\noindent
where $p_i''= d x_i/dt''$, such that the dynamics only depends on the rescaled energy $E=H/B \chi$.
The following study employs the Hamiltonian \ref{eq:hamFPU3} and we 
omit  the primes in order to simplify the notation.

We begin our exploration of the system's dynamics by addressing first its static properties regarding its equilibria, i.e. the
 roots of the $N$-dimensional gradient (critical points) of the potential 
\begin{equation}
\label{potentialV1}
V= \sum_{i=1}^{N} \bigg[\sin x_i \sin x_{i+1}
-2 \cos x_i \cos x_{i+1} \bigg],
\end{equation}
\noindent
given by the system of equations ($\forall i= 1,2,\ldots, N$)
\begin{equation}
\label{deriV1}
\frac{\partial V}{\partial \theta_i}= \bigg[\cos x_i \sin x_{i+1} +2 \sin x_i \cos x_{i+1}
+\cos x_i \sin x_{i-1} +2 \sin x_i \cos x_{i-1} \bigg]=0.
\end{equation}

\subsection{Equilibrium points}
From the inspection of Eqs.~\ref{deriV1}, we find the following critical points summarized in Fig.~\ref{fi:eq_points}:

\medskip
\begin{itemize}
\item[(i)] The head-tail configuration of the dipoles $\{x_i=0, \forall i\}$ or $\{x_i=\pi, \forall i\}$  (Fig.~\ref{fi:eq_points} (i)).
This critical point is a minimum of the
potential~\ref{potentialV1} (see Appendix A) with energy 
$E_m=-2 N.$ For the sake of simplicity, in the following we shift all the energies of the system  by $2 N$, such
that this
minimum energy becomes zero, i.e. $E_m=0$.

\item[(ii)] The tail-tail and head-head configurations with alternating angles
$0$ and $\pi$, $\{x_i=\pi \left[1 \pm(-1)^i\right]/2, \forall i\}$ (Fig.~\ref{fi:eq_points} (ii)).
 These critical points are degenerate maxima of the
potential~\ref{potentialV1} (see Appendix A) with energy $E_M= 4 N$ (shifted by  $2 N$). 

\item[(iii)] Configurations of alternating $2b$ blocks of an arbitrary number of dipoles $n_i$ ($i=1,2,\ldots,2b$)
where within each block $i$ all $n_i$ dipoles  are either oriented as
$x_k=\pi$ or $x_k=0$. The potential energy of this configuration is $E_m$ plus the
potential energy excess of all pairs of dipoles left and right to the interfaces
of two neighboring blocks with oppositely aligned dipoles. For our PBC this adds up to
the total energy  $E_{s1}=8 b$  (shifted by  $2 N$).
An example of such a configuration is shown in Fig. \ref{fi:eq_points} (iii) where, taking
into account the PBC, there are six blocks of 
dipoles with alternating polarization resulting in a total energy $E_{s1}=24$.

These critical points are argued to be saddle points of $\textrm{rank}=2 b$ in the Appendix A.
In particular, for the maximum number of possible blocks,
$2b=N$,  we recover the configuration of maximum potential energy, 
$E_M=4 N$, which is indeed a critical point of $\textrm{rank}=N$.
It is worth noting that all these saddle points are highly degenerate with respect to the length of the blocks and both 
their energy and their rank (number of negative eigenvalues of the Hessian), depend only on the number
of blocks $2b$ and not on the number of dipoles $n_i$ within each block.

\begin{figure}[t]
\centerline{\includegraphics[scale=0.5]{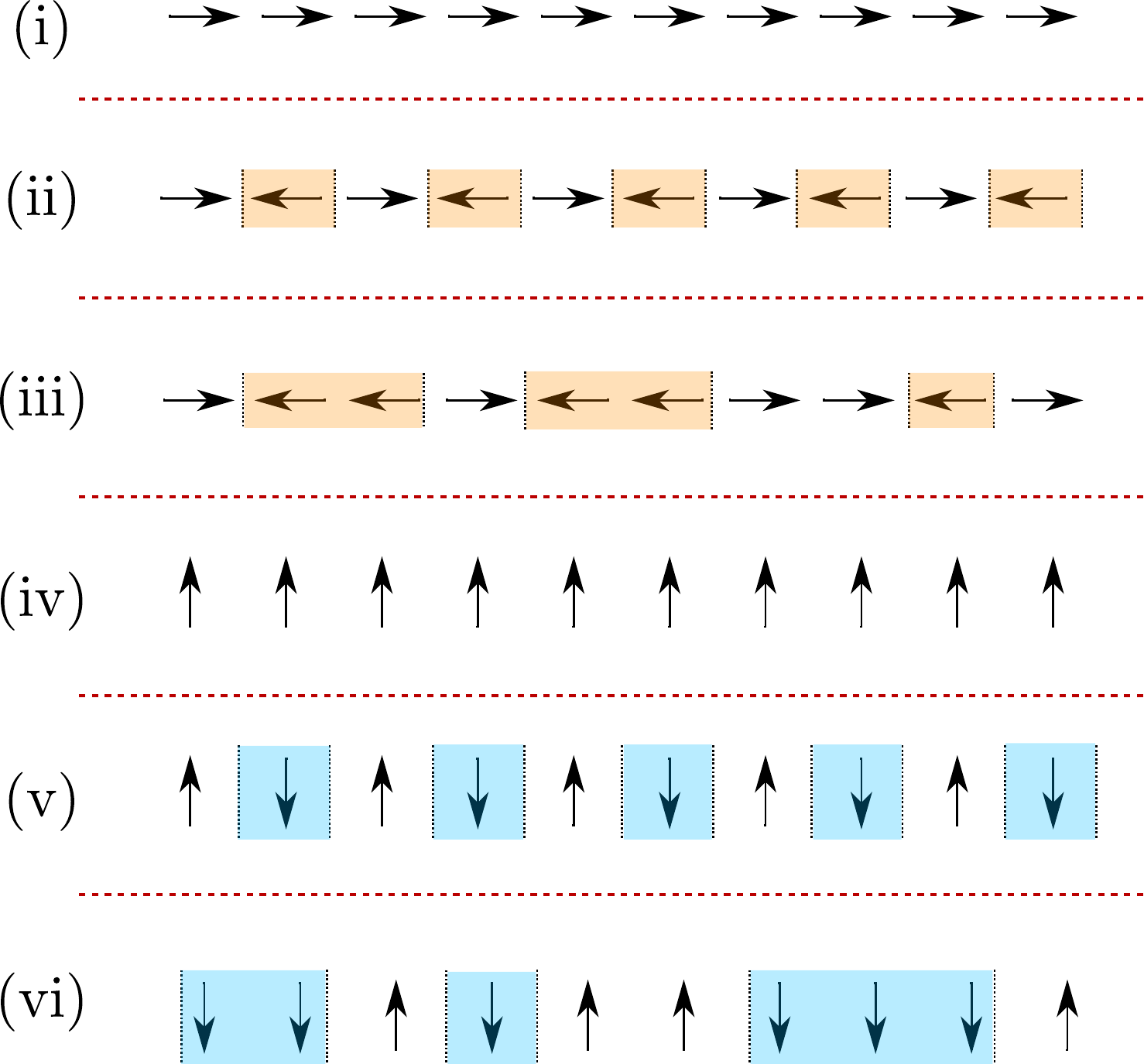}}
\caption{Schematic representation of the six families of equilibrium points (i)-(vi) discussed in the main text
in terms of the angle $x_i$.}
\label{fi:eq_points}
\end{figure}

\item[(iv)] The two configurations with  $\{x_i=\pi/2, \forall i\}$ or $\{x_i=-\pi/2, \forall i\}$
(Fig.~\ref{fi:eq_points} (iv)) which are
saddle points (see Appendix A) with  energy
$E_{s2}= 3 N$ (shifted by  $2 N$).

\item[(v)] The configurations with alternating $\pi/2$ and $-\pi/2$, $\{x_i= \pm(-1)^i\pi/2, \forall i\}$
(Fig.~\ref{fi:eq_points} (v)) which represent
saddle points (see Appendix A) with  energy $E_{s3}= N$ (shifted by  $2 N$).


\item[(vi)] Configurations of alternating $2b$ blocks of an arbitrary number of dipoles $n_i$ ($i=1,2,\ldots,2b$)
where within each block $i$ all $n_i$ dipoles  are either oriented as
$x_k=\pi/2$ or $x_k=-\pi/2$. Following the same discussion as in the equilibrium configuration (iii), the potential energy of this configuration is $E_{s2}=3N$ [case (iv)] minus the
potential energy excess of all pairs of dipoles left and right to the interfaces
of two neighboring blocks with opposite up and down aligned dipoles, such that the
total energy is $E_{s4}=3 N -4b$  (shifted by  $2 N$).
An example of this configuration is shown in Fig. \ref{fi:eq_points} (vi) where, taking
into account the PBC, there are six blocks of 
dipoles with alternating up and down orientation resulting in a total energy $E_{s4}=18$.

\end{itemize}

The above six families of critical points (Fig.~\ref{fi:eq_points}) allow one to get a glimpse of the high complexity
of the landscape of the $N$-dimensional potential energy surface $V$ [see Eq.~\ref{potentialV1}].
The discussed energy hierarchy of these families should be reflected in the 
dynamics of the dipole chain. In particular, we expect 
a linear dynamics for small excitations around the potential minimum  $E_m=0$ and a quite regular behavior 
for total excitation energies $E$ below the energy of the lowest saddle point, i.e. for $E<\min\left(E_{s1}\right)=8$. However, for $E>8$, and due to the larger accessible phase space regions which involve
also different equilibria, we expect to encounter a nonlinear behavior.
It is worth noting that for values of $b$ close to one, the energy $E_{s1}=8 b$
of the corresponding saddle points is much smaller than the maximum energy $E_M=4 N$ of the
potential $V$. Hence,  one should expect nonlinear behavior even for small excitation energies 
$E\gtrsim\min\left(E_{s1}\right)=8$.

In the following, we present results for the dynamics of the dipole chain for different excitation energies, spanning the  three aforementioned regions
with qualitatively different dynamical behavior, i.e. the linear ($E\ll 8$), the regular ($E \lesssim 8$) and the
irregular ($E \gtrsim 8$) regime.

\section{The linear behavior}
\label{sec:linear0}
The equations of motion of the Hamiltonian \ref{eq:hamFPU3}
can be written as:
\begin{equation}
\label{ecumovi}
\ddot x_i = -(\cos x_i \sin x_{i+1} + 2 \sin x_i \cos x_{i+1}+\cos x_i \sin x_{i-1} + 2 \sin x_i \cos x_{i-1})
\end{equation}
For low-energy excitations, e.g., small oscillations around the head-tail equilibrium configuration
$\{x_i=0, \forall i\}$ or $\{x_i=\pi, \forall i\}$ of minimum energy $E_m$, the linear approximation of the equations of motion \ref{ecumovi} yields
\begin{equation}
\label{linear}
\ddot x_n = -(x_{n-1} + 4 x_n+ x_{n+1} ),\quad n=1,...,N.
\end{equation}
As it is well-known, the system of linear differential equations~\ref{linear} can be solved in terms of $N$ normal modes 
$(Q_k, P_k)$ \cite{cross1955},
\begin{eqnarray}
\label{fourier}
Q_k(t) &=& \frac{1}{\sqrt{ N}} \sum_{n=1}^{N} x_n(t) \exp \left(i \frac{2 \pi k n}{N}\right),\quad k=1,...,N 
\nonumber \\
& & \\
P_k(t) &=& \frac{1}{\sqrt{ N}} \sum_{n=1}^{N} p_n(t) \exp \left(i \frac{2 \pi k n}{N} \right),\quad k=1,...,N,
\nonumber
\end{eqnarray}
\noindent
where $p_n(t)=\dot x_n(t)$.

\begin{figure}[t]
\centerline{\includegraphics[scale=.4]{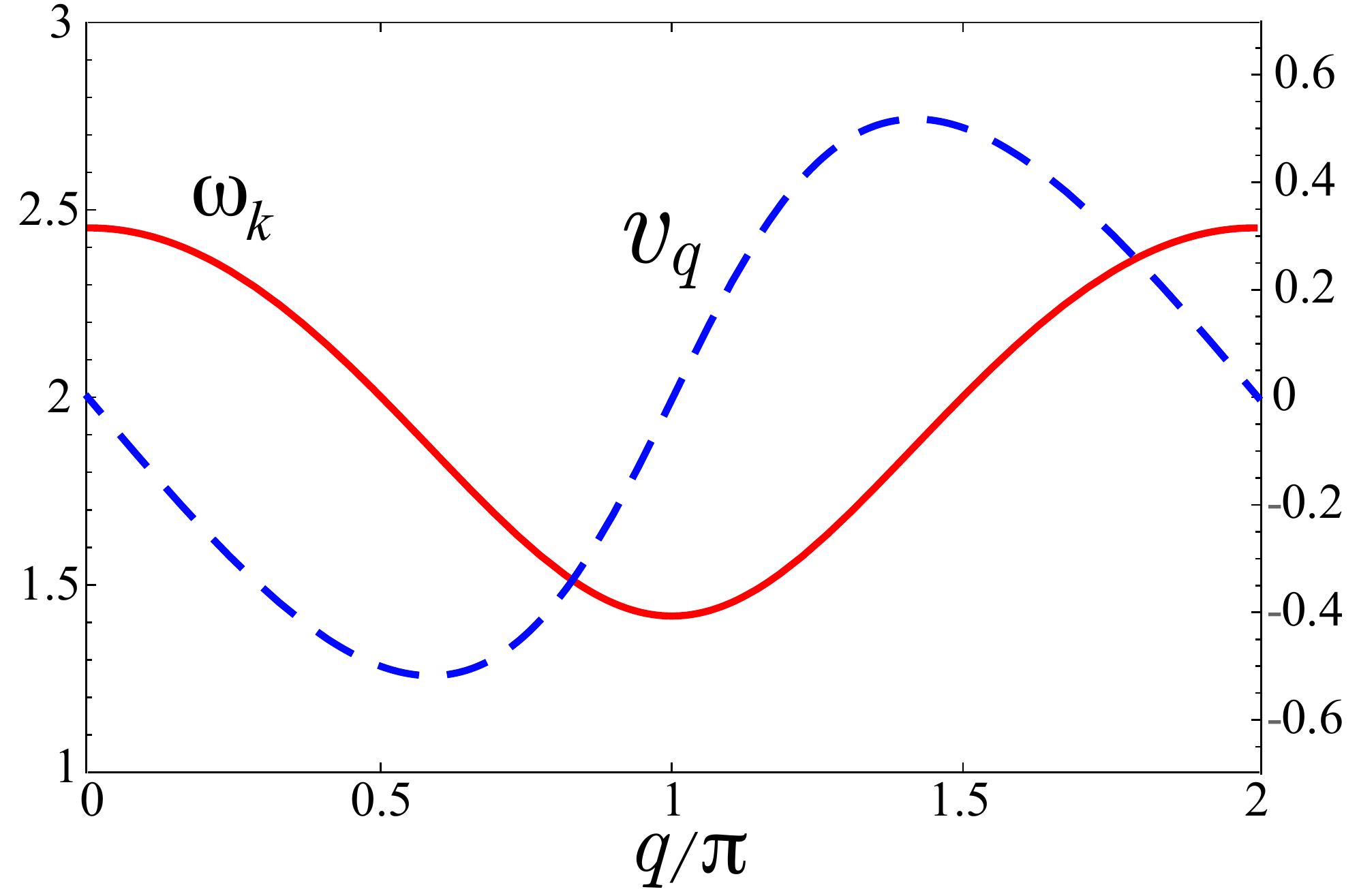}}
\caption{Dispersion relation $\omega_k$ (red solid curve) and  group velocity $v_q$ (blue shaded curve)
as a function of $q/\pi$ with  $q=2\pi k/N$ being the wave number.}
\label{fi:dispersion}
\end{figure}

In the normal mode variables $(Q_k, P_k)$, the Hamiltonian $H_0$ associated
to the linear system~\ref{linear} reads 
\begin{equation}
\label{linearModes}
H_0= \sum_{k=1}^{N} {\cal E}_k=E_0,\qquad {\cal E}_k = \frac{1}{2} \left( |P_k|^2+ \omega_k^2 \ |Q_k|^2\right),
\quad k=1,...,N
\end{equation}
where $\omega_k$ and ${\cal E}_k$ are the
frequency and  the (harmonic) energy of each normal mode, respectively. The sum of the energies of all normal modes $\{{\cal E}_k\}$
yields the total harmonic energy $E_0$, corresponding   to the Hamiltonian $H_0$ of the linearized system~\ref{linear}.
The frequency $\omega_k$ relates to the wave number $k$ through the dispersion relation
\begin{equation}
\label{dispersion}
\omega_k = \sqrt{4 + 2 \cos q}, \quad q = \frac{2 \pi k }{N}
\end{equation}
derived from Eqs.~\ref{linear}-\ref{fourier}. The above expression (Eq.~\ref {dispersion}) has already been deduced
in e.g. the study of molecular chains \cite{Ratner1} and has the form depicted in Fig.~\ref{fi:dispersion}(red solid line).
As expected, the frequency $\omega_k$ is  2$\pi$-periodic with $q$ and it enjoys a
reflection symmetry with respect to $q=0$ and $\pi$. 
As we can observe in Fig.~\ref{fi:dispersion}(solid red line),  the
linear spectrum is optic-like with the frequency $\omega_k$
possessing an upper bound (maximum)  $\omega_k=\sqrt{6}$ for $q \rightarrow 0$ (long-wavelength limit)
and a lower bound (minimum) $\omega_k=\sqrt{2}$ for $q=\pi$ (short wavelength limit).
\noindent
From the dispersion relation \ref{dispersion}, the
group velocity $v_q$ of the normal modes can be derived
\begin{equation}
\label{velocity_group}
v_q = \frac{d \omega_k}{d q} = -\frac{\sin q}{\omega_k}.
\end{equation}
As  shown in Fig.~\ref{fi:dispersion} (blue shaded line), $v_q$ vanishes when $\omega_k$ reaches its maximum or
minimum value, indicating
that the normal modes with the longest and the shortest wavelengths are non--propagating
modes. In contrast,
for $q/\pi =\arccos(-2+\sqrt{3})/\pi \approx 0.59$ and $q/\pi =1+ \arccos(2-\sqrt{3})/\pi \approx 1.41$ the group velocity reaches its maximum amplitude $|v_q|\approx 0.52$, rendering the corresponding normal modes the fastest propagating 
ones in the system.

In the current study we are interested in the time  propagation of single dipole excitations through the dipole chain
for different values of the excitation energy. More specifically, starting from the head-tail 
configuration (Fig.~\ref{fi:eq_points}(i)) of minimal
energy $E_m=0$, we excite at $t=0$ a single dipole, supplying it with an excess energy
$\Delta K$. In all our calculations we  use a chain of 200 dipoles with PBC, a fact that allows us to
excite a specific dipole
(here the $100$th) without loss of generality.
  The initial conditions $(x_i(0), p_i(0))$
of our system at  $t=0$ are given therefore by
\begin{eqnarray}
\label{iniC}
x_i(0)&=&p_i(0)=0, \qquad \mbox{for} \quad  i \ne 100,\nonumber\\[2ex]
\Delta K &=& \frac{p_{100}(0)^2}{2} + 4 [1-\cos x_{100}(0)].
\end{eqnarray}
\noindent
Using these initial conditions, we investigate
the time propagation of the excitation  by integrating numerically the equations of motion  \ref{ecumovi}
for the dipole chain.
In order to achieve a high accuracy,  we integrate Eqs. \ref{ecumovi} using an
explicit Dormant--Prince Runge--Kutta
algorithm of eighth order with step size control and dense output~\cite{hairer}. The results of these
integrations are subsequently compared to those extracted by a symplectic and
symmetric Gauss method of six stages~\cite{hairer2}.  Up to the same prescribed error tolerances, in
all cases the numerical results obtained with both methods are the same.

During the integration  we record at  each time step, besides the phase space variables $x_i(t)$ and $p_i(t)$ of each dipole, 
also  the harmonic energy contribution
${\cal E}_k(t)=\frac{1}{2} \left( |P_k(t)|^2+ \omega_k^2 \ |Q_k(t)|^2\right)$ 
of each Fourier mode $(Q_k(t),P_k(t))$
resulting from the Fourier transform (Eq.~\ref{fourier}) of the numerically extracted $\{x_i(t), p_i(t)\}$. 
We emphasize here that all these quantities are recorded for the exact
equations of motion (Eq.~\ref{ecumovi}) of our system
and not for their linearized form (Eq. \ref{linear}) discussed above.

As we have briefly  mentioned in the previous section, for very low values of the excitation energy 
$\Delta K \ll 8$ (much lower than the energy of the first saddle
point) we expect a linear behavior of the dipole chain, with the Eqs. \ref{linear}
describing appropriately the small oscillations of the dipoles around the head-tail equilibrium configuration.
In this linear regime, 
the harmonic energy ${\cal E}_k(t)$ 
stored in each Fourier mode $(Q_k(t),P_k(t))$ remains almost constant in time,
since the Fourier modes are the approximate (uncoupled) normal modes of the
system, and therefore the total excitation energy $\Delta K$ is roughly equal 
to the total harmonic energy $E_0(t)=\sum_{k=1}^{N} {\cal E}_k(t)$ distributed among the $N$ Fourier modes  of the system [see Eq.~\ref{linearModes}].

For larger excitation energies $\Delta K$ the behavior of the system is expected to be in general  nonlinear, involving  a transfer of energy  between 
the different Fourier modes  $(Q_k(t),P_k(t))$ due to their coupling. The higher the degree of such a
nonlinear mode-coupling, the higher
we expect to be the deviation of $\Delta K$ (the total energy of our system, involving all the couplings between the modes) from the total harmonic
energy contribution $E_0(t)$ resulting from the  modes $(Q_k(t),P_k(t))$ which are assumed to be uncoupled. Therefore, we can use this deviation between $\Delta K$ and $E_0(t)$
as an indicator of the degree of nonlinearity in the system. 

In particular, we define the function $C_1(\Delta K)$
\begin{equation}
\label{correC1}
C_1(\Delta K)=\frac{<{E_0}>}{\Delta K}, \quad \left<E_0\right>= \frac{1}{t_f} \int_{0}^{t_f} E_0(t)\ dt,\end{equation}
where $\left<E_0\right>$ is the time average of the total harmonic energy $E_0(t)$ of the Fourier modes $\{Q_k(t),P_k(t)\}$ up to a (large) final time $t_f$. 
According to our above discussion  $C_1(\Delta K)=1$ for a linear system, where the Fourier modes, coinciding with its normal modes, are  uncoupled.
The closer the function $C_1(\Delta K)$  is to $1$,
the closer the exact dynamics of the system is expected to be to linear.

\begin{figure}[t]
\centerline{\includegraphics[scale=1]{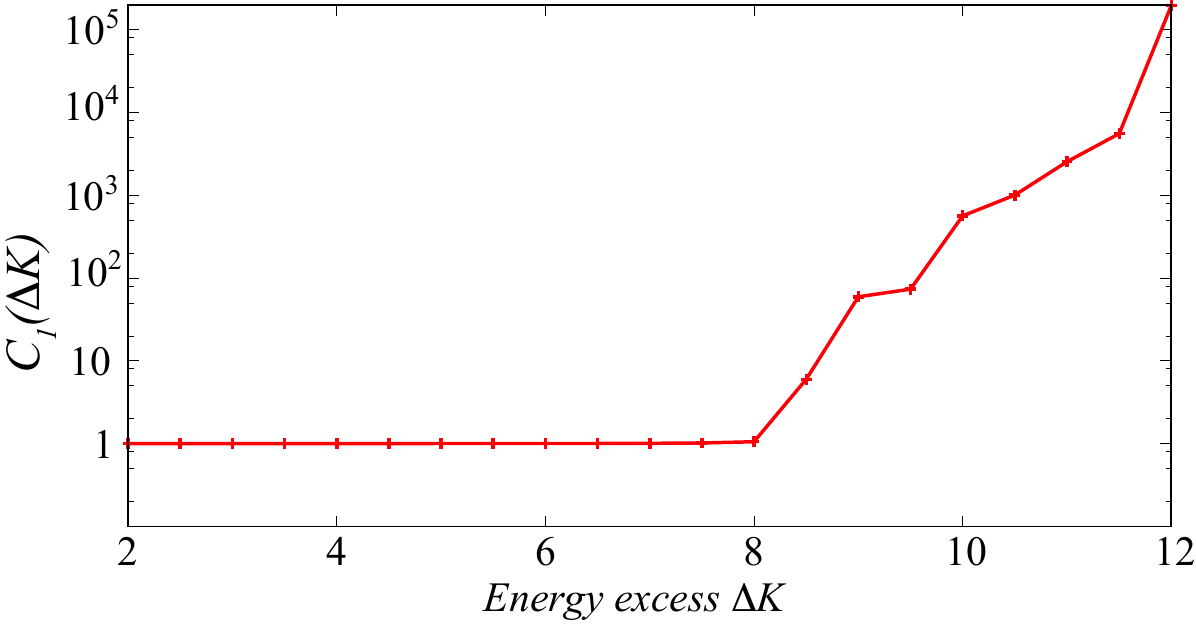}}
\caption{
Dependence of  $C_1(\Delta K)$ [see Eq.~\ref{correC1}] on the excitation energy $\Delta K$ for a single initially
excited dipole (Eq.~\ref{iniC}). Each point of the curve  corresponds to
the average over an ensemble of 40 simulations with the same initial excess energy but for different initial conditions.
Note that a logarithmic scale scale is used for $C_1(\Delta K)$.}
\label{fi:meanC1}
\end{figure}

We present in Fig.~\ref{fi:meanC1} the behavior of $C_1(\Delta K)$ for 
excess energies $\Delta K \in [2, 12]$. Apart from the time average in the definition of 
$C_1(\Delta K)$ (Eq.~\ref{correC1}), we have also performed for each point of Fig.~\ref{fi:meanC1}  an average over $40$ different sets of initial conditions
(different choices of $x_{100}(0)$ and $p_{100}(0)$, all corresponding to the same excess energy $\Delta K$, see
Eqs. ~\ref{iniC}).
We observe that for $\Delta K < 8$ the dynamics of the system is only weakly nonlinear, since
the average of the total harmonic energy contribution $\left<E_0\right>$ of
the Fourier modes is a good approximation
to the total energy $\Delta K$ of the system ($C_1(\Delta K)\approx1$).
In this region we expect that the linear
normal modes couple only weakly, leading to  minor energy transfer between different modes, 
but keeping the corresponding harmonic total energy $E_0(t)$ approximately constant, equal to $\Delta K$.

In contrast, for excess energies $\Delta K > 8$ the value of $C_1(\Delta K)$ increases rapidly, with
the average total harmonic energy  $\left<E_0\right>$ of the Fourier modes obtaining much larger values than the total excitation energy $\Delta K$,
a fact that indicates a highly nonlinear behavior. Although the value $\Delta K = 8$ cannot be considered as a precise threshold between the regimes of a
weakly  and a highly nonlinear behavior, this value can be perceived as an upper bound, above  which  the system reacts to localized energy excitations
in a highly nonlinear way. It is worth noting that this upper bound
($\Delta K \approx 8$)
coincides with the energy $E_{s1}=8$ of the lowest saddle point consisting of two blocks of dipoles with opposite polarization. 
After overcoming the energetic barrier of the first saddle point,
the available phase space  of the
system increases dramatically, offering  possibilities for various dynamical behaviors. Interestingly, we see
that the total harmonic energy contribution $E_0$ of the Fourier modes  is
always larger than the total energy $E\equiv \Delta K$ of the system.
In other words, the contribution of the coupling between the Fourier modes to the energy, representing
the nonlinear interaction, is negative, e.g., it is attractive.

This observation can be justified by the $4$th-order expansion of 
the total potential $V$ (Eq.~\ref{potentialV1}) around the equilibrium position $\{x_i=0, \forall i\}$. Such an expansion yields
\begin{equation}
\label{expansionV1}
V \approx V_1 =\sum_{i=1}^N \left(x_i^2 + x_{i+1}^2 + x_i x_{i+1}\right)
-\sum_{i=1}^N \left(\frac{(x_i^2+x_{i+1}^2) (x_i+x_{i+1})^2}{12} + \frac{x_i^2 \ x_{i+1}^2 }{3}\right),
\end{equation} 
where the negative energy contribution of the nonlinear terms to the total potential energy is clearly observed.
The potential $V_1$ resembles the Fermi-Pasta-Ulam~${\beta}$-model
(FPU-${\beta}$) with a potential of the form
$V_{FPU-\beta}=\sum_{i=1}^N\left[\frac{1}{2}{\left(x_i- 
x_{i+1}\right)^2}+\frac{\beta}{4} \left(x_i- x_{i+1} \right)^4\right]$, 
since both contain only quartic nonlinear terms. However, in our model, in contrast to the FPU models, the
 degree of nonlinearity is fixed and cannot be controlled 
by varying the system parameters (such as $\beta$). Besides, while the linear spectrum of our
problem is optic-like (Fig. \ref{fi:dispersion}), the FPU models display an acoustic-like
linear spectrum~\cite{A867,A837} with no frequency gap for long wavelengths $q \rightarrow 0$.

\section{The Weakly Nonlinear Regime}
\label{sec:linear}
In this section we present in detail the response of the dipole chain to single
local perturbations in the weakly nonlinear 
regime of small excitation energies ($\Delta K < 8$).
Following a similar scheme as in Sec.~\ref{sec:linear0}, given a chain of 200 dipoles with PBC in the head-tail
configuration of minimal energy $E_m=0$, we locally excite  at $t=0$ the 100th dipole of the chain
by supplying it with an excess of kinetic energy $\Delta K$. Thus, the initial conditions $(x_i(0), p_i(0))$ of our
system at $t=0$ read
\begin{eqnarray}
\label{iniC2}
x_i(0)&=&p_i(0)=0, \qquad \mbox{for} \quad  i \ne 100,\nonumber\\[2ex]
x_{100}(0)&=&0, \quad p_{100}(0)=\sqrt{2\Delta K}.
\end{eqnarray}
\noindent
In order to study the propagation of this initially localized excitation along the dipole chain
 we calculate numerically the time evolution of the local energies  $E_k(t)$
\begin{eqnarray}
\label{local}
E_k(t)&=& \frac{p_k(t)^2}{2}+
\frac{1}{2} \bigg[\sin x_k(t) \sin x_{k+1}(t)
-2 \cos x_k(t) \cos x_{k+1}(t) + \\ \nonumber
&&\sin x_k(t) \sin x_{k-1}(t)
-2 \cos x_k(t) \cos x_{k-1}(t)\bigg],
\end{eqnarray}
\noindent
which indicate the amount of energy stored in each dipole in relation to its nearest neighbours.

\begin{figure}[t]
\centerline{
\includegraphics[scale=0.7]{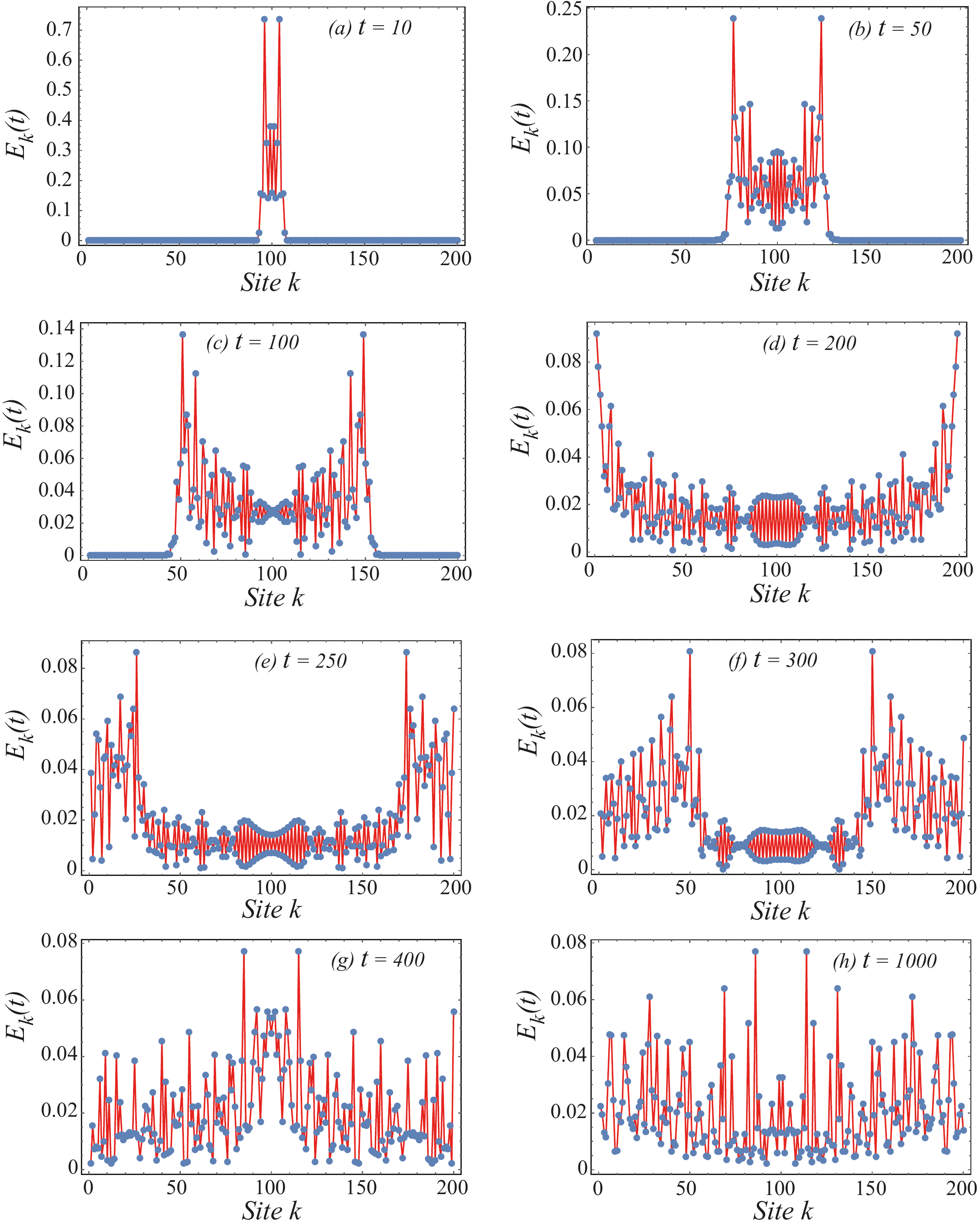} 
}
\caption{Spatial (site) distribution of the local energies $E_k(t)$ for  different time instants.
The number of dipoles in the chain is $N=$200 and the initial kinetic energy
excess provided to the central dipole at site 100 is $\Delta K=4$.}
\label{fi:spatial_distribution}
\end{figure}

\begin{figure}[t]
\centerline{
\includegraphics[scale=1]{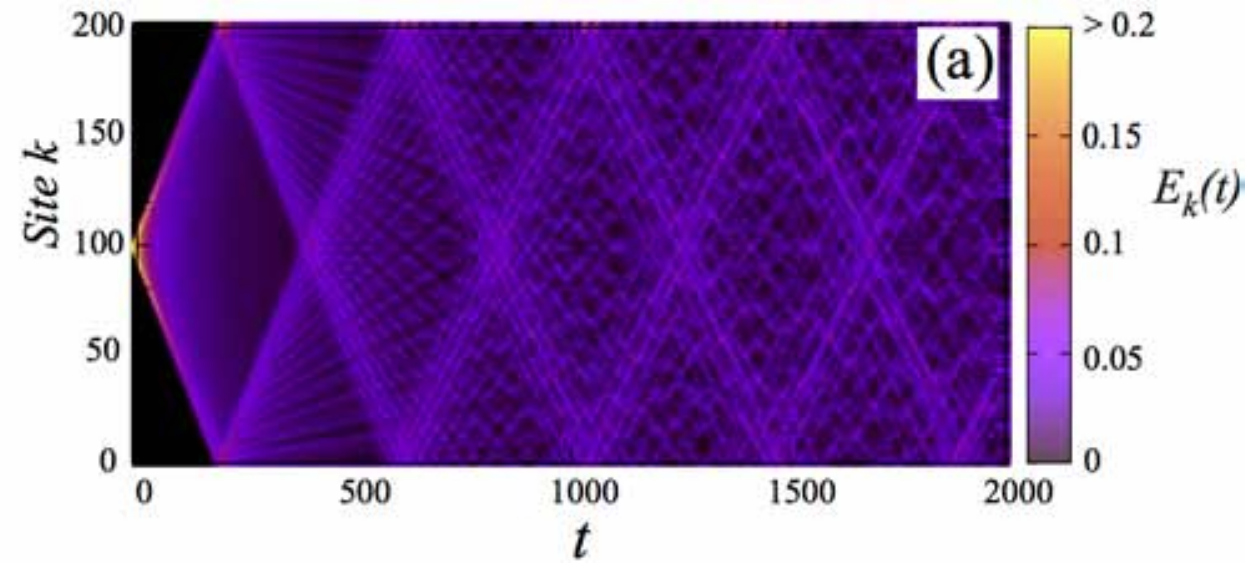}}

\bigskip
\centerline{\includegraphics[scale=0.5]{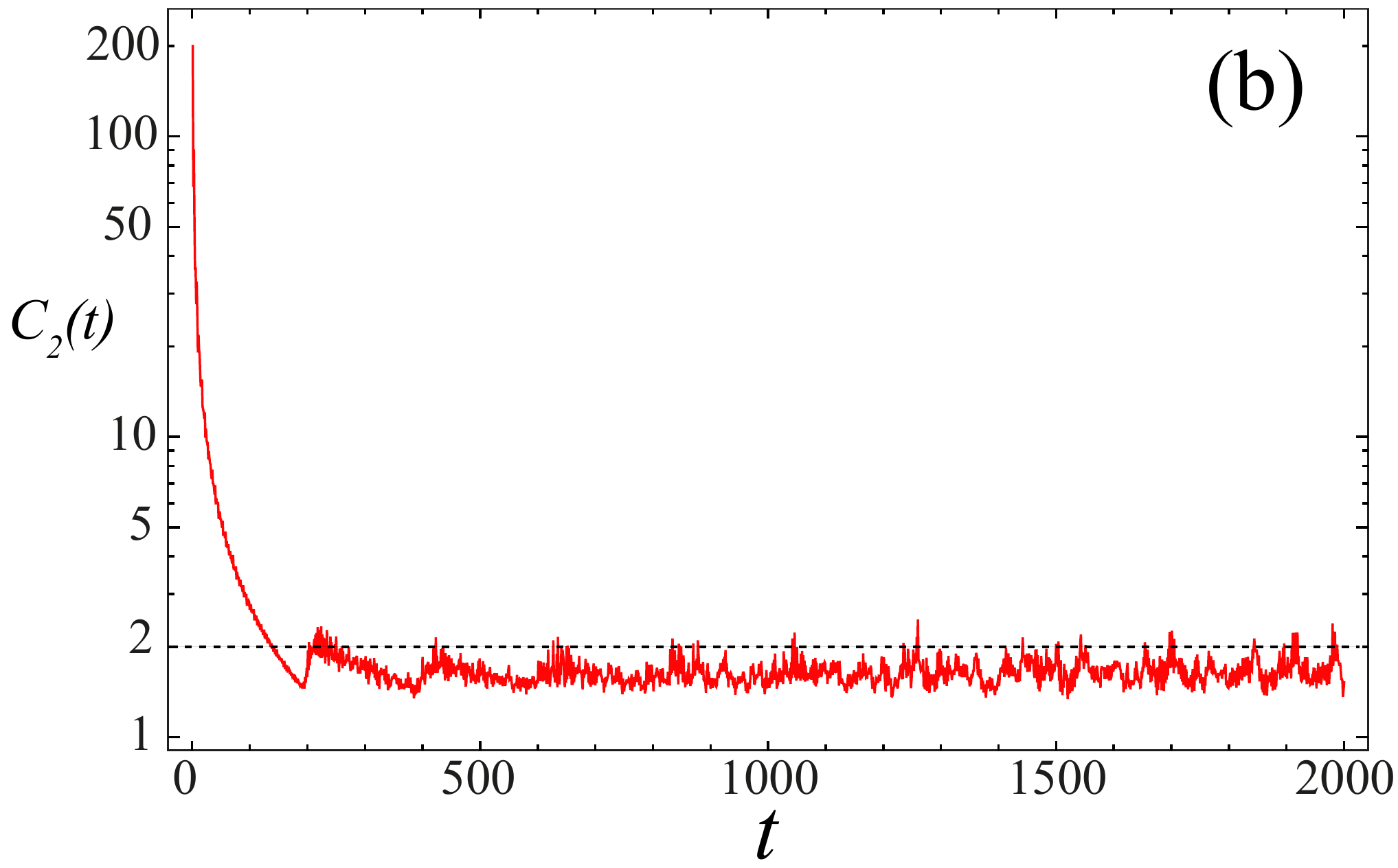}}
\caption{a) Time evolution of the
local energy  $E_k(t)$ of the dipoles (depicted by color).
The number of dipoles in the chain is $N=$200 and the initial kinetic energy
excess provided to the central dipole at site 100 is $\Delta K=4$.
b) Time evolution of  the inverse participation ratio $C_2(t)$  (Eq.~\ref{correC2}) for $N=$200 and $\Delta K=4$.
Note that a logarithmic scale is used for $C_2(t)$.}
\label{fi:colormap1}
\end{figure}

The local energy profiles for $\Delta K=4$ at different time instants shown in Fig. \ref{fi:spatial_distribution}
provide a glimpse of the different steps of the excitation propagation. Shortly after the excitation,
 most of the excess energy is transfered to the nearest neighbors of the initially excited dipole ($100$th), which become the main energy carriers 
initiating the energy spreading along the chain.
Indeed, at short times $t=10, 50$ and $100$ [see Figs.\ref{fi:spatial_distribution}(a)-(c)], the
excitation transfer  is clearly induced by two (symmetric) energy fronts
that propagate along the chain.
At  $t\approx 200$, the energy excitation reaches the ends of the chain [see
Fig.\ref{fi:spatial_distribution}(d)] having transfered an amount of energy to every dipole in the chain, causing their oscillations.
This  yields a propagation velocity $v_p\approx 0.5$, close to the maximum value of the group velocity  $|v_q|\approx 0.52$  found for the 
linear case (see Eq.~\ref{velocity_group}).

Due to the PBC of the system for
$t \gtrsim 200$ the excitation  continues its propagation from the outer dipoles (at sites 1, 200) to the inner ones (located at sites around 100),
i.e. the direction of propagation is reversed such that 
for $t \approx 400$ the excitation reaches again the central dipoles of the chain [see
Figs.\ref{fi:spatial_distribution}(e)-(g) corresponding to $t=250, 300$ and $400$].
As the chain is progressively excited, the sharp intensity peaks of the propagation fronts observed at short times (Fig.\ref{fi:spatial_distribution}(a))
decays  significantly (Figs.\ref{fi:spatial_distribution}(e)), indicating that the system tends to thermalize, reaching for long times
energy equipartition [see Fig. \ref{fi:spatial_distribution}(h) for $t=1000$].

According to the above discussion, a global picture of the time evolution of the
local energy $E_k(t)$ is given in terms of a  color map in
Fig.~\ref{fi:colormap1} (a). After the $100$th dipole is 
excited with an excess energy $\Delta K=4$, the excitation energy is gradually distributed along the
chain by means of the two aforementioned 
symmetric energy fronts. As the system approaches the energy equipartition state, the 
 energy fronts are distorted and their intensity decreases.

\medskip
To quantify the localization of the energy along the chain, we use the following function, usually
termed as the inverse participation ratio \cite{A851}
\begin{equation}
\label{correC2}
C_2(t)= N \frac{\sum_{k=1}^N E_k(t)^2}{\Delta K^2},
\end{equation}
with $E_k$  being the local energies given by Eq.~\ref{local}.
When the  excitation is maximally localized, i.e. the total excitation energy energy $\Delta K$ of the system is carried by a single dipole, the
value of $C_2$ is $N$, while if there is
complete equipartition ($E_k \approx \Delta K/N ~\forall k$) $C_2=1$. The time evolution  of $C_2$  is shown in 
Fig.~\ref{fi:colormap1} (b). Starting from a fully localized excitation ($C_2(0)=200$) the excitation propagation quickly leads to a regime where the excitation energy 
is almost equally partitioned among all the dipoles ($C_2\lesssim 2$). 

\section{The highly NonLinear Regime}
Following the same scheme as in Sec.~\ref{sec:linear}, we excite at $t=0$  the 100th dipole from the head-tail ground state of a 200-dipole chain,
supplying it with a kinetic energy excess $\Delta K>8$.

\begin{figure}[t]
\centerline{
\includegraphics[scale=1]{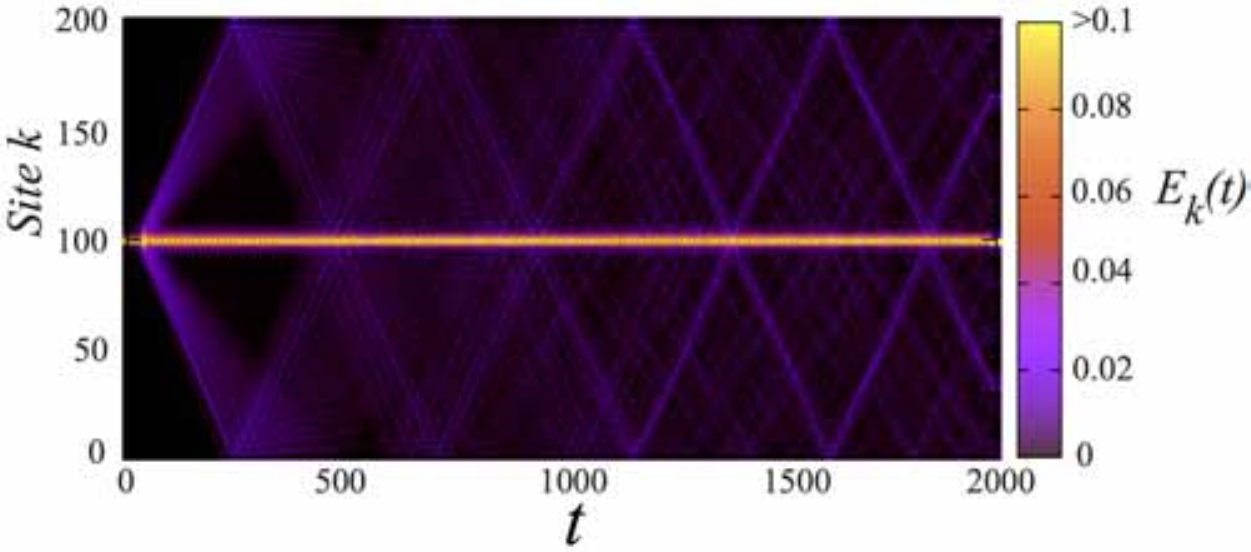}}
\caption{Time evolution of the
local energy  $E_k(t)$ of the dipoles (depicted by color).
The number of dipoles in the chain is $N=$200 and the initial kinetic energy
excess given to the central dipole 100 is $\Delta K=12$.}
\label{fi:colormap3}
\end{figure}

\begin{figure}[t]
\centerline{
\includegraphics[scale=0.7]{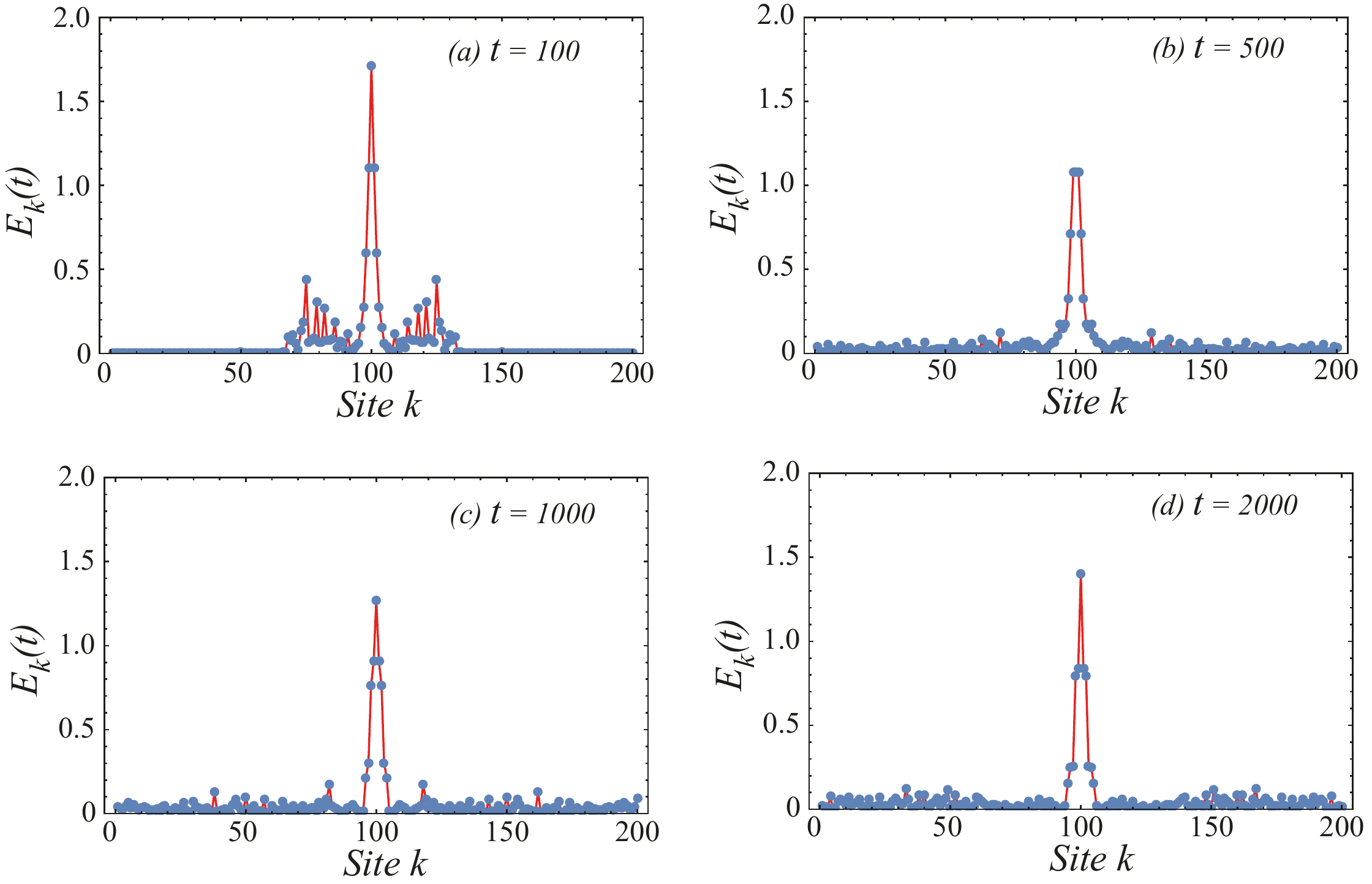} }
\caption{Spatial (site) distribution of the local energies $E_k(t)$  for different time instants.
The number of dipoles in the chain is $N=$200 and the initial kinetic energy
excess given to the central dipole at site 100 is $\Delta K=12$.}
\label{fi:spatial_distribution2}
\end{figure}

A typical propagation scheme in this highly nonlinear  regime is the one obtained for an excess energy $\Delta K=12$.
 For this value the time evolution of the spatial distribution of the
local energies $E_k(t), ~k=1,2,\ldots 200$ is depicted in Fig.~\ref{fi:colormap3}. We observe 
 a robust excitation around the $100$th dipole, indicating that the system does not reach energy
 equipartition up to long times. 
Indeed, the initially excited dipole $100$
 shares predominantly energy with a few of its neighbors so that a significant part of the excess energy remains 
localized around it. 

This fact is emphasized  in
Fig.~\ref{fi:spatial_distribution2} where the local energy profiles for
$t=$100, 500, 1000 and 2000 are depicted.
At short times (see Fig.~\ref{fi:spatial_distribution2}(a) for $t=$100)
a small propagation front emerges 
whose energy after some time is distributed among all the dipoles of the chain
(see Fig.~\ref{fi:spatial_distribution2}(b) for $t=$500).
However, the excitation energy
of the few central dipoles (close to the initially excited one) remains much larger than that of the other outer dipoles,
creating overall a highly localized profile which persists in time (see Fig.~\ref{fi:spatial_distribution2}(c)-(d)).
\begin{figure}[t]
\centerline{
\includegraphics[scale=0.5]{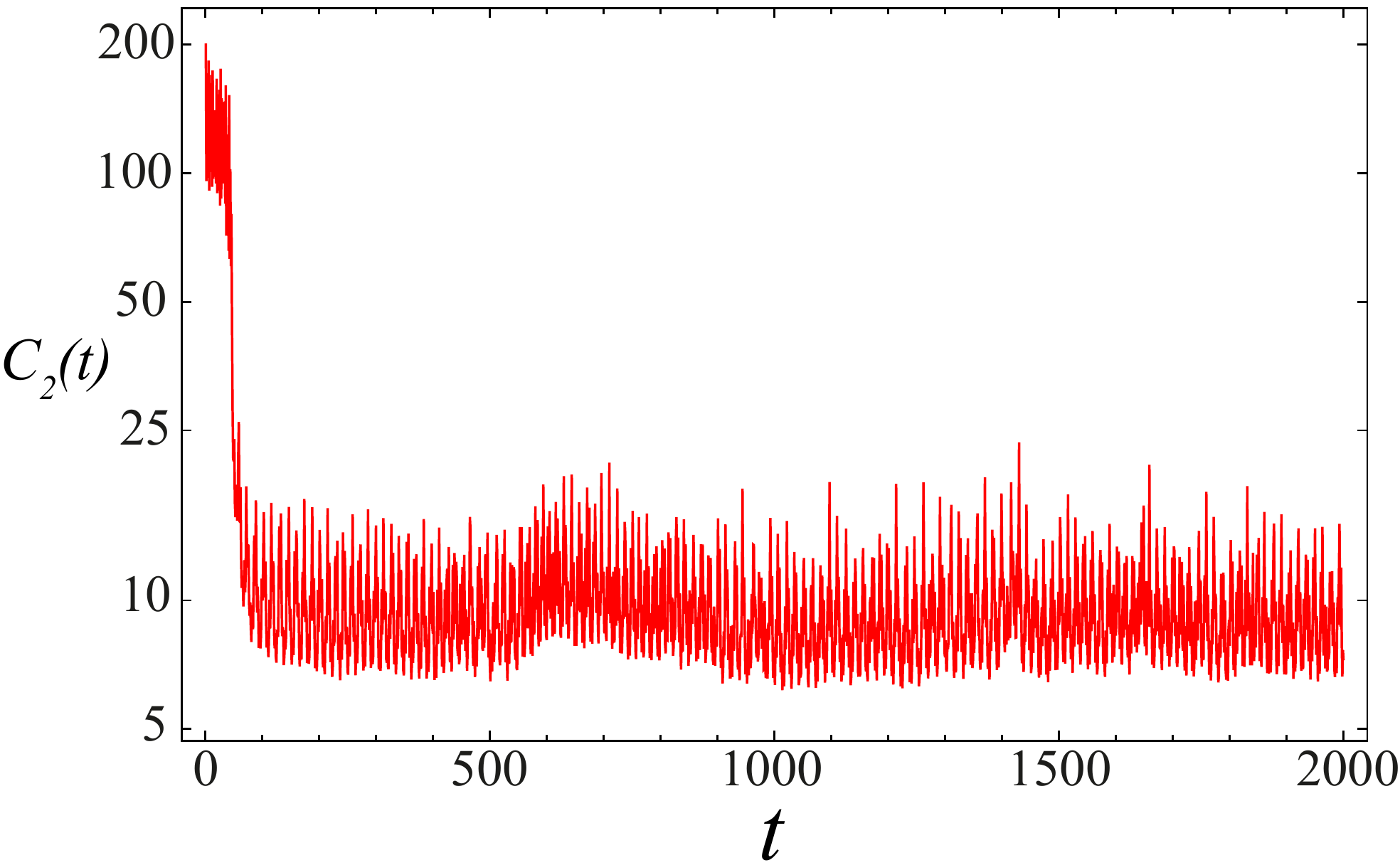}}
\caption{Time evolution of the inverse  participation ratio $C_2(t)$ for
 a kinetic energy excess  $\Delta K=12$.}
\label{fi:correC2B1DK12CI1}
\end{figure}

As in the previous section, the degree of localization of this local excitation with an energy excess
$\Delta K=12$ can be quantified 
by means of the inverse participation ratio $C_2$ (Eq.~\ref{correC2}). In its time evolution, shown in
Fig.~\ref{fi:correC2B1DK12CI1}, we observe a rapid decrease followed by asymptotic
high-amplitude oscillations around $C_2 \approx 10$. 
This asymptotic value of $C_2$ is an order of magnitude larger than the corresponding value of $C_2$ in the weakly nonlinear regime (see Fig.~\ref{fi:colormap1} (b)), which
points to the much stronger localization of the excitation.
This high degree of localization suggests the existence of a strong nonlinearity since, for the linear case 
the dispersion of the excitation energy along the complete chain dominates the dynamics, leading to an approximately
equipartition regime in terms of local energies $E_k$.

\begin{figure}[t]
\centerline{
\includegraphics[scale=0.5]{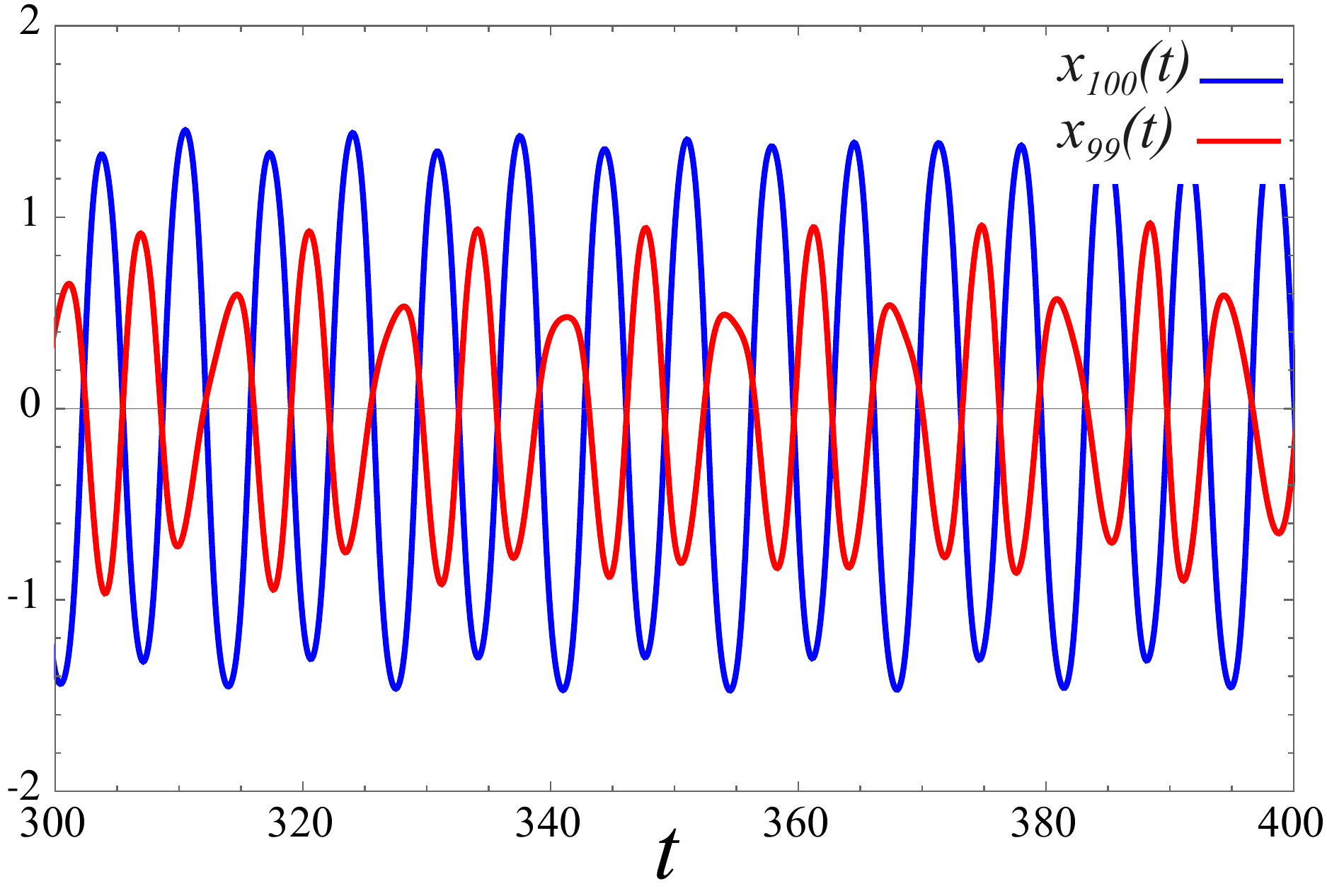}}
\caption{Time evolution of the angles $x_{99}(t)$
and $x_{100}(t)$ of the dipoles 99 and 100 in the interval $300 \le t \le 400$.
Excess energy is $\Delta K=12$.}
\label{fi:excitationX100DK12}
\end{figure}

\begin{figure}[t]
\centerline{
\includegraphics[scale=0.4]{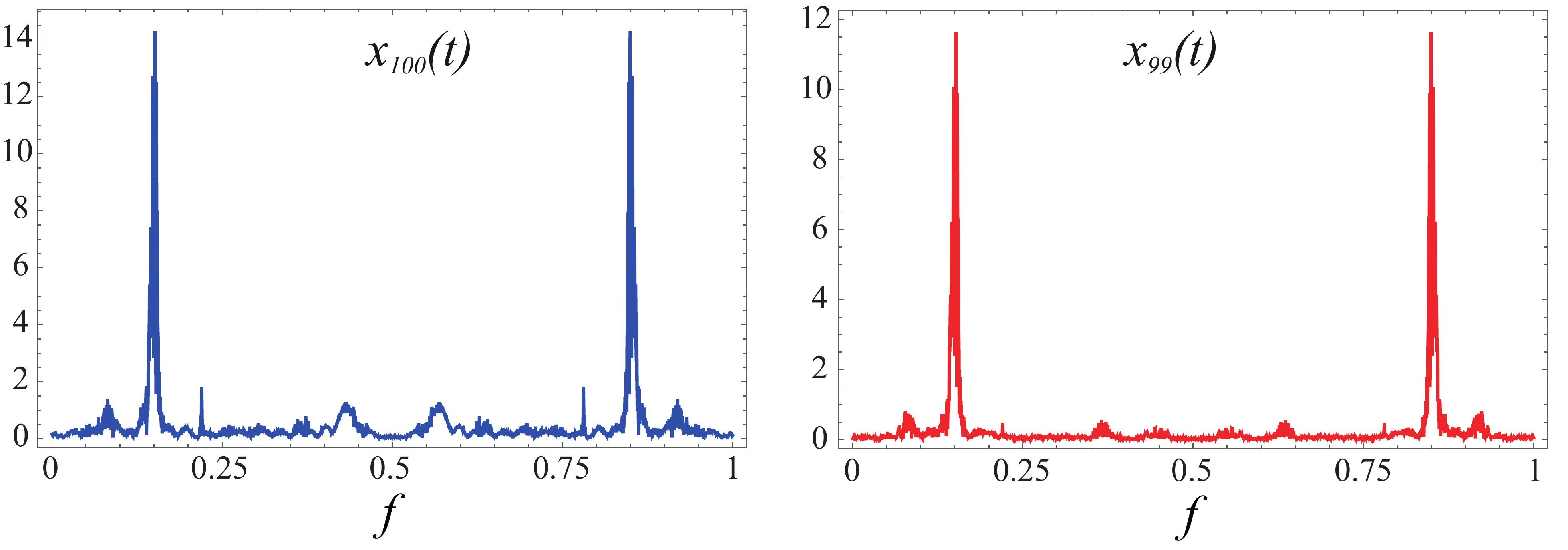}}
\caption{Fourier spectra of $x_{99}(t)$
and $x_{100}(t)$ of the dipoles 99 and 100.
The excess energy $\Delta K=12$ is provided to the 100th dipole.}
\label{fi:fourierSpectra}
\end{figure}

Even more, it turns out that the localization of the excitation energy in
Figs.~\ref{fi:colormap3}-\ref{fi:spatial_distribution2} can be linked to discrete breather solutions
of the nonlinear equations of motion of our system (Eq. \ref{ecumovi}). 
Briefly speaking, a discrete breather is a  spatially localized  {\sl exact} periodic solution of the nonlinear equations
of motion of a given discrete lattice (for more details, we refer the reader to \cite{A867,A868}).
The non-resonant condition between the frequency $\Omega$ of
a breather solution and the dispersion relation $\omega_k$ prevents the existence of breathers with a
frequency in the
linear spectrum, such that the breather frequency $\Omega$ should always lie outside the linear spectrum $\omega_k$.
In our system, for $\Delta K=12$, we have seen  (Fig.~\ref{fi:spatial_distribution2}) that
a major part of the excitation energy $\Delta K$ 
stored initially on the $100$th dipole 
remains  localized in the few dipoles surrounding it up to long times. Moreover, when the
time evolution of the angles $x_{99}(t)$
and $x_{100}(t)$ of the dipoles 99 and 100 respectively are examined (see Fig.\ref{fi:excitationX100DK12}), we
find that their motion is in both cases fairly periodic (oscillatory) with a period of  $\tau\approx 6.7$. 
The periodicity of these
oscillations is justified by  Fig. \ref{fi:fourierSpectra}, where the
Fourier spectra of $x_{99}(t)$ and $x_{100}(t)$ are depicted.
Indeed, we observe that these 
spectra   exhibit  a strong peak at a frequency $f\approx 0.15$
(with its symmetric counterpart at $f \approx 0.85$),
reflecting the fact that the corresponding signals can be approximated as oscillations with a single frequency
$\Omega=2\pi f \approx 2 \pi \times 0.15 \approx 0.94$, corresponding
to a period of  $\tau \approx 6.67$. Since this oscillation frequency, $\Omega \approx 0.94$, is outside
(more precisely below) the linear spectrum depicted in Fig. \ref{fi:dispersion} (red line), we have
strong evidence that the localized excitation of the dipole chain observed  in Figs.~\ref{fi:colormap3}-\ref{fi:spatial_distribution2}
during its time evolution corresponds to a discrete breather.
\begin{figure}[t]
\centerline{\includegraphics[scale=0.75]{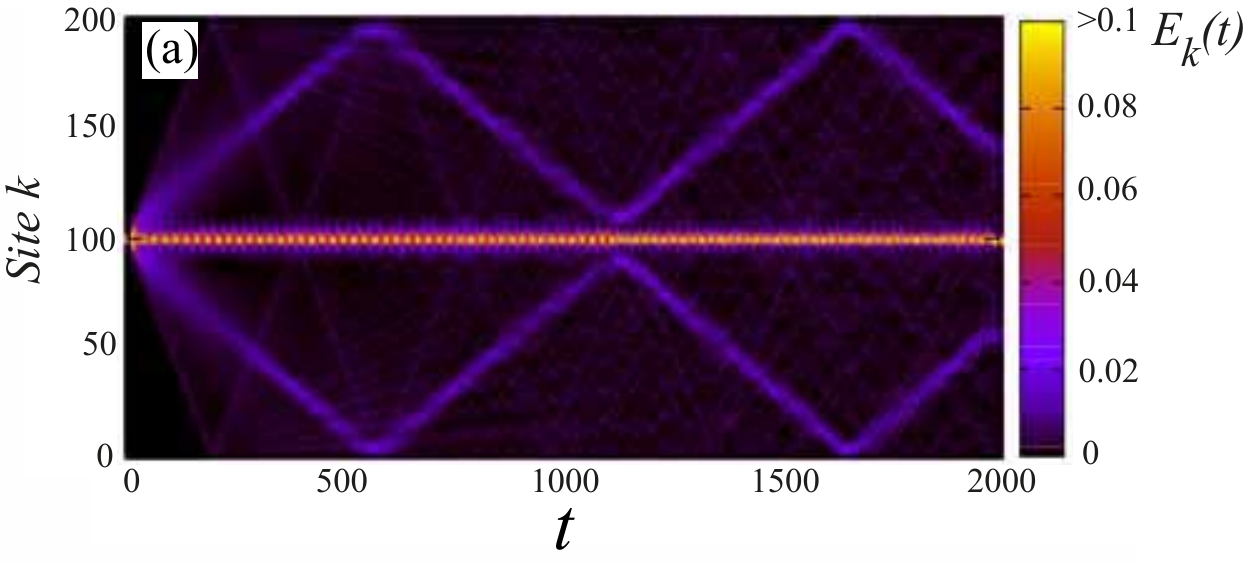}
\includegraphics[scale=0.75]{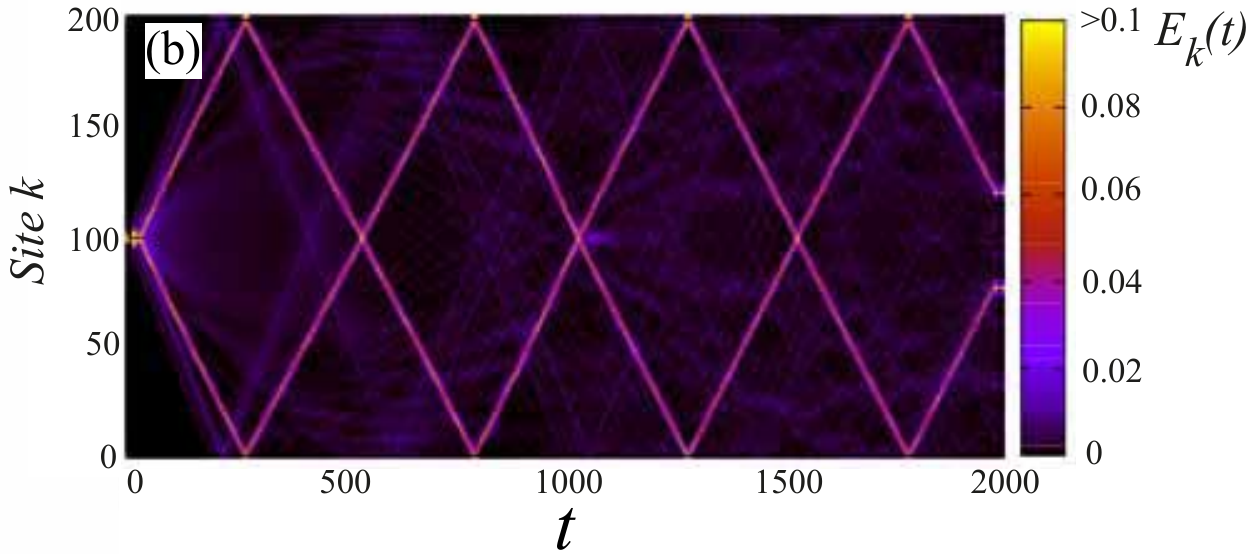}}

\centerline{\includegraphics[scale=0.75]{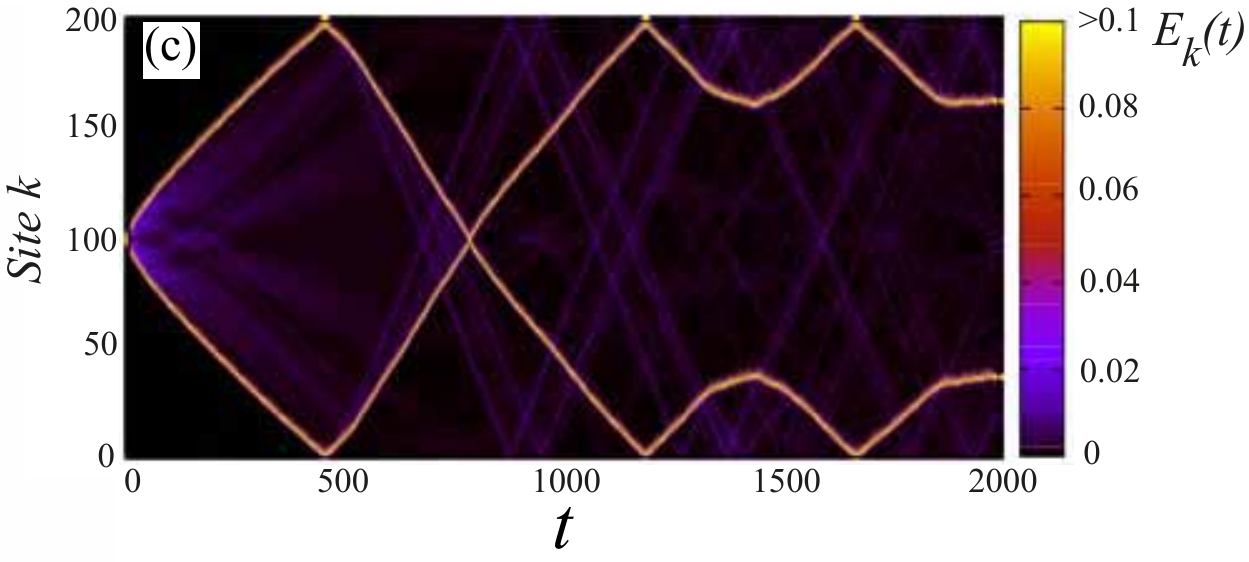}
\includegraphics[scale=0.75]{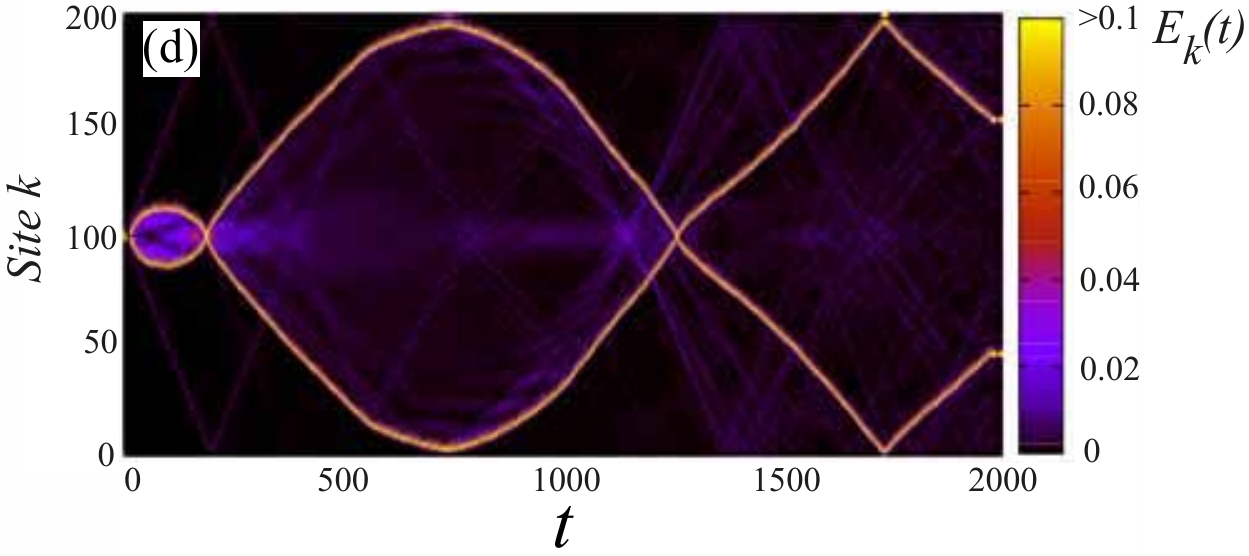}}
\caption{Time evolution of the
local energy  $E_k(t)$ of the dipoles (depicted by color) for  four values of the
excess energy (a) $\Delta K$=10.776, (b) $\Delta K$=11.128, (c) $\Delta K$=10.2
and (d) $\Delta K$=11.12.}
\label{fi:colormap4}
\end{figure}

Besides the typical discrete breather pattern of energy localization shown in Fig.~\ref{fi:colormap3}, the dipole chain exhibits additional propagation schemes,
depending on the value of the initial excitation energy $\Delta K$, all of them involving a high degree  of energy
localization. The time evolutions for the local energies $E_k(t)$ for four  cases
$\Delta K$ = 10.776, 11.128, 10.2 and 11.12, 
corresponding to different propagation patterns, are shown in Figs.~\ref{fi:colormap4}(a)-(d).
The main difference between these
is the behavior of the principal energy carriers, i.e. the dynamics of those sites that carry the largest amount of excitation energy.
For $\Delta K$ = 10.776 and 11.128
[Fig.~\ref{fi:colormap4}(a)-(b)], the energy of the system is highly localized in two
energy carriers that follow trajectories of a regular periodic character.
In contrast, for $\Delta K$ = 10.2 and 11.12 [Fig.~\ref{fi:colormap4}(c)-(d)] the principal energy
carriers follow rather complex trajectories
which, given their strong localization and complexity, could 
be linked to the so-called chaotic breathers \cite{A851}. Contrary to the concept of a discrete breather as a localized excitation which is 
a solution of the nonlinear equations
of motion of the lattice, a
chaotic breather is an excitation of chaotic nature that may appear as a 
response to initial local excitations of the lattice, a situation 
that, as it will be argued below, bears strong similarities to the one described here 
 regarding the propagation of a localized excitation in our dipole chain in the highly 
nonlinear regime ($\Delta K \gtrsim 8$).

\begin{figure}[t]
\centerline{\includegraphics[scale=0.75]{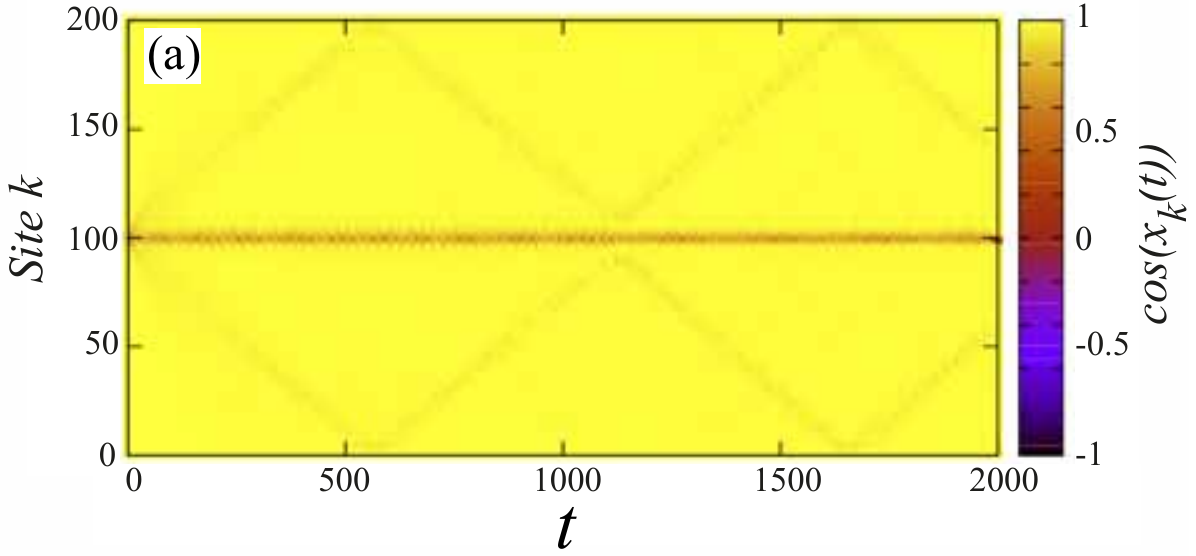} \quad
\includegraphics[scale=0.75]{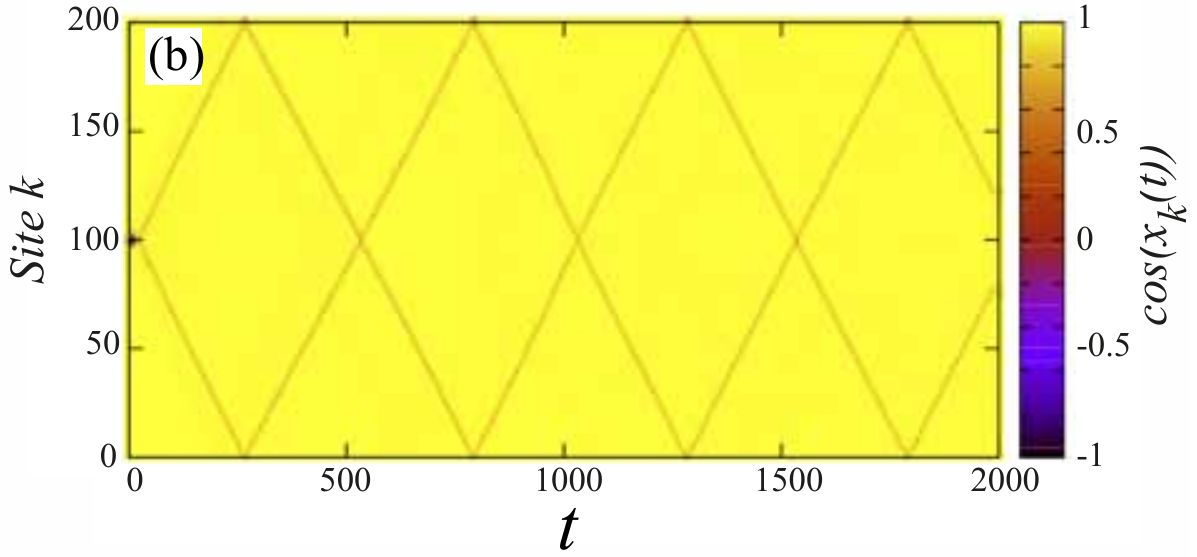}}

\centerline{\includegraphics[scale=0.75]{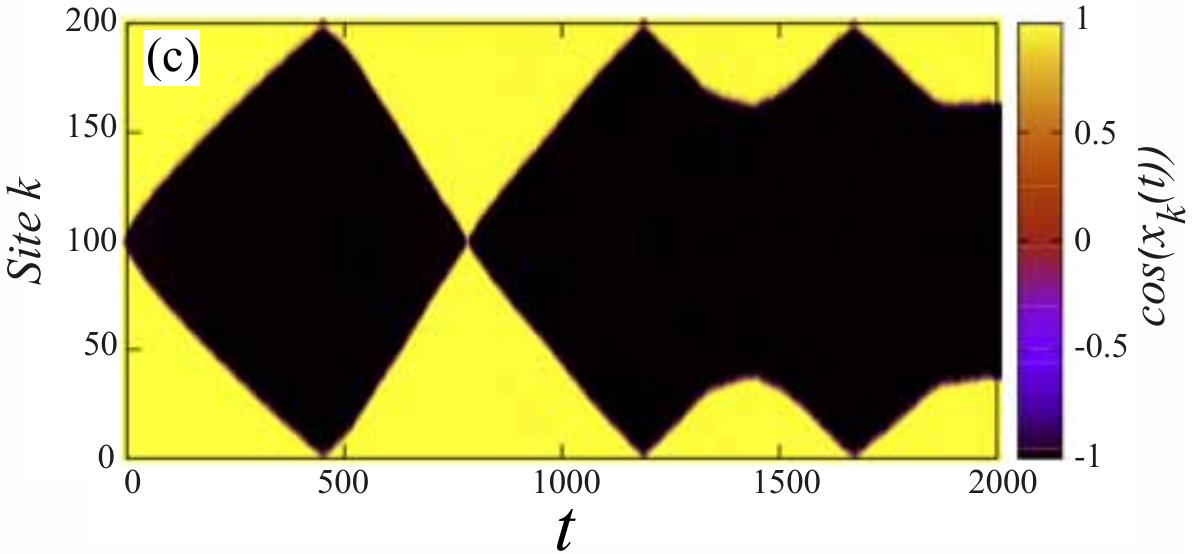} \quad
\includegraphics[scale=0.75]{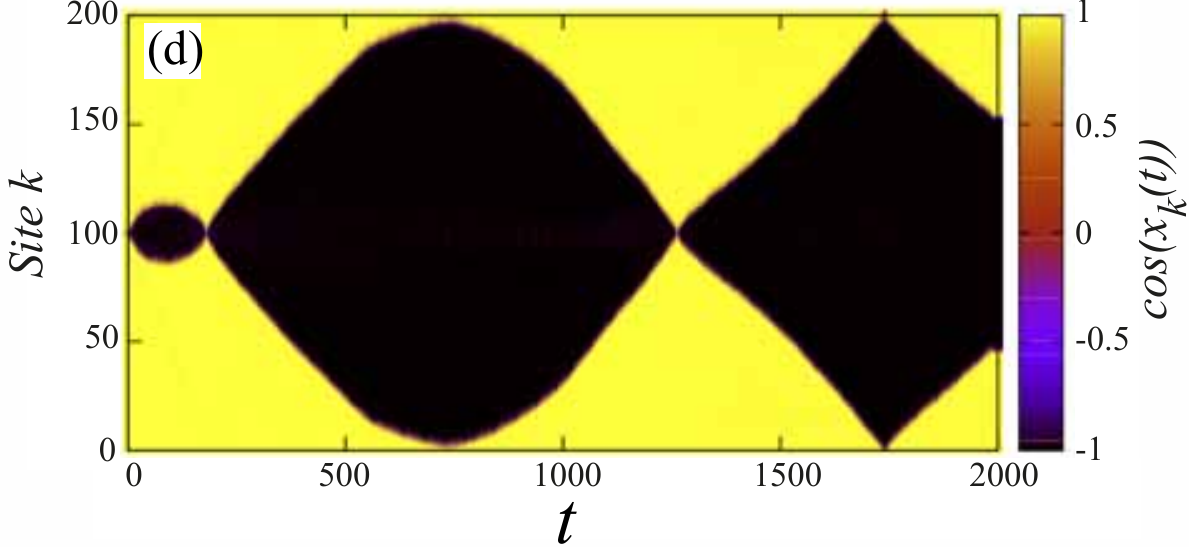}}
\caption{The time evolution of  $\cos(x_k)$ (depicted by color) for different values of the excess energy $\Delta K$: (a) $\Delta K$=10.776, (b)$\Delta K$=11.128, (c)$\Delta K$=10.2
and (d) $\Delta K$=11.12. as in Fig.\ref{fi:colormap4}.}
\label{fi:cosf1}
\end{figure}

Apart from the different patterns of energy propagation observed in Fig.~\ref{fi:colormap4}
for the different values of $\Delta K$, it turns out that also the configurations $\{x_k\}$
evolve differently in time. We illustrate this fact in Fig.~\ref{fi:cosf1}
where the time evolution of  $\cos(x_k(t))$ is shown for the same excitation energies, 
$\Delta K =$10.776, 11.128, 10.2 and 11.12, as those
considered in Fig.~\ref{fi:colormap4}. For  $\Delta K = 10.776$ and $11.128$ 
(see Figs.~\ref{fi:cosf1}(a)-(b)) we observe the expected behavior: 
except for the energy carrying rotors (ECRs),
which exhibit fast long-amplitude oscillations (fast changing $x_k$) while propagating along the chain
 leading to the corresponding traces in Figs.~\ref{fi:cosf1}(a)-(b), 
all the remaining dipoles are mainly polarized
in the same direction $\{x_k=0\}$ ($\cos(x_k)=1$)
as in the ground state configuration.

In contrast, for $\Delta K = 10.2$ and $11.12$ (Fig.~\ref{fi:cosf1}(c)-(d)) the situation
is dramatically different. Instead of a single polarized region ($\cos (x_k)=1$), like the 
yellow background in Figs.~\ref{fi:cosf1}(a)-(b), two regions of opposite polarization 
($\cos (x_k)=1$ and $\cos (x_l)=-1$ )
emerge during the time evolution. These regimes of locked phases ($\{x_k=0\}$ and $\{x_l=\pi\}$ respectively) corresponding to domains of opposite polarization
appear spontaneously and they are
dynamically separated by two propagating domain walls provided by the two fast rotating ECRs (compare Figs. \ref{fi:cosf1} (c),(d)  with  
Figs. \ref{fi:colormap4} (c),(d).
 In particular, in the course of the dynamics, the dipoles lying between the two ECRs spontaneously flip, 
 forming a domain of opposite polarization  ( $\cos (x_l)=-1$, black region in Figs. \ref{fi:cosf1} (c),(d)) compared to that of the
 ground state (yellow region in  Figs. \ref{fi:cosf1} (c),(d)).

Although the origin of this spontaneous phase locking 
is not entirely clear, it can be related to the existence of the lower
saddle point equilibria of energy $E_{s1}= 8<\Delta K$, ($b=1$, Sec.II.A. (iii) ).
This assumption relies on the resemblance of the topology between the phase locked states and the
highly degenerate
saddle point equilibrium configurations consisting of two blocks: one  with a given number $n$
of dipoles with $x_l=\pi$ and another with the remaining $N-n$ dipoles polarized along $x_k=0$. 
As mentioned in Sec. II, all such saddle points, consisting of two domains of opposite polarization  are highly degenerate,
since their total potential energy $E_{s1}=8$ depends only on the number of domain walls (here two) and not 
on the number of dipoles on each domain ($n$ and $N-n$ respectively). With  an excitation energy $\Delta K=12>E_{s1}$ 
a spontaneous dynamical transition from the fully polarized ground state to the first saddle point is energetically  possible and
therefore can occur for certain initial conditions (Figs.~\ref{fi:cosf1}(c)-(d)). During the time evolution of such a state the domain walls, identified with the ECRs, shift (Figs.~\ref{fi:cosf1}(c)-(d)),
 following the complex trajectories shown in Figs.~\ref{fi:colormap4}(c)-(d), a process that due to the aforementioned degeneracy of the
first saddle point does not cost any  energy.

It is worth noticing that phase locked states with more than two domains
(more than two domain walls) never appear in our simulations
considering an excitation energy $\Delta K \in [4 , 12]$, since already the energy of the
second saddle point, $E_{s1}= 16$, 
consisting of four domains (four domain walls), is inaccessible. It should be
observed, however, for $\Delta K>16$.

 A closer look at Fig.~\ref{fi:colormap4} and Fig.\ref{fi:cosf1}, particularly a comparison between
 Fig.~\ref{fi:colormap4}(b) and Fig.~\ref{fi:colormap4}(d) 
(also between Figs.~\ref{fi:cosf1}(b) and Figs.~\ref{fi:cosf1}(d)),  
leads to the conclusion that even a tiny change of the excitation energy $\Delta K$ (here only by $0.07\%$) can lead to a completely different propagation and configuration pattern.
We have checked  that this is the case also when a infinitesimal perturbation is added to the initial values of the phase space variables ($x_i(0), p_i(0)$). This strong sensitivity to the initial conditions  is
the hallmark of the chaotic nature of our system in the region
of excitation energies $\Delta K>8$.

As a measure of this sensitivity to the initial conditions  in the dipole chain
(i.e. its degree of chaoticity), we use the method of the
Orthogonal Fast Lyapunov Indicators (OFLI).
In a nutshell, given an {\sl m}-dimensional flow defined by
\begin{equation}
\label{flow}
\frac{d {\bf r}}{dt} = {\bf f}({\bf r},t),
\end{equation}
we examine the time evolution of the variational vector $\delta{\bf r}(t)$ given by the
(first) variational equations
\begin{equation}
\label{Dflow}
\frac{d \delta{\bf r}}{dt} = \frac{\partial{\bf f}({\bf r},t)}{\partial {\bf r}} \delta{\bf r}.
\end{equation}
For  given initial conditions ${\bf r}(0)$ and  $\delta{\bf r}(0)$, the numerical integration of the systems 
of differential equations \ref{flow} and \ref{Dflow} up to a given
final time $t_f$ allows the definition  of the OFLI as follows~\cite{ofli2,ofli22,A709}
\begin{equation}
\label{def:ofli}
\mbox{OFLI}({\bf r}(0), \delta{\bf r}(0), t_f) =
\sup_{0 \le t \le t_f} \log || \delta{\bf r}(t)^\bot||,
\end{equation}
where $\bot$ indicates the orthogonal component to the flow
of the variational vector $\delta{\bf r}$. The main advantage of the OFLI is that it
provides computationally cheap information about the degree of
regularity/chaoticity of a given orbit.
In particular,  $\delta{\bf r}(t)^\bot$ increases
linearly with time for regular resonant orbits and  exponentially for chaotic ones ~\cite{ofli2,ofli22,A709},
attaining therefore for long times $t_f$ much larger values for chaotic orbits than the ones for regular orbits.
For near-integrable Hamiltonian systems, a rigorous proof of
this behavior can be found in \cite{A709}.
We note that with the formulation \ref{def:ofli}, there is a dependence of the value of the
OFLI on the initial conditions of the
variational vector $\delta{\bf r}(0)$. In order to get rid of this dependence we follow the steps found in~\cite{barrio,barrio1},
incorporating also  the second order variational equations in the computation of the indicator.
For our dipole chain, we have
calculated, as a function of the energy excess $\Delta K$, the
OFLI for  trajectories as those examined so far, featuring initially a single dipole excitation
with initial conditions ${\bf r}(0)=\{x_k(0)=0~\forall k, p_{100}(0)=\sqrt{2 \Delta K}, p_k(0)=0 ~\forall k\ne100\}$.
In our calculations, we
stop the computation of the OFLI either when it reaches the cutoff value 9, marking  a chaotic trajectory, or when the 
computation time exceeds our final time $t_f=5000$, selected empirically according to many numerical simulations.

\begin{figure}[t]
\centerline{\includegraphics[scale=0.8]{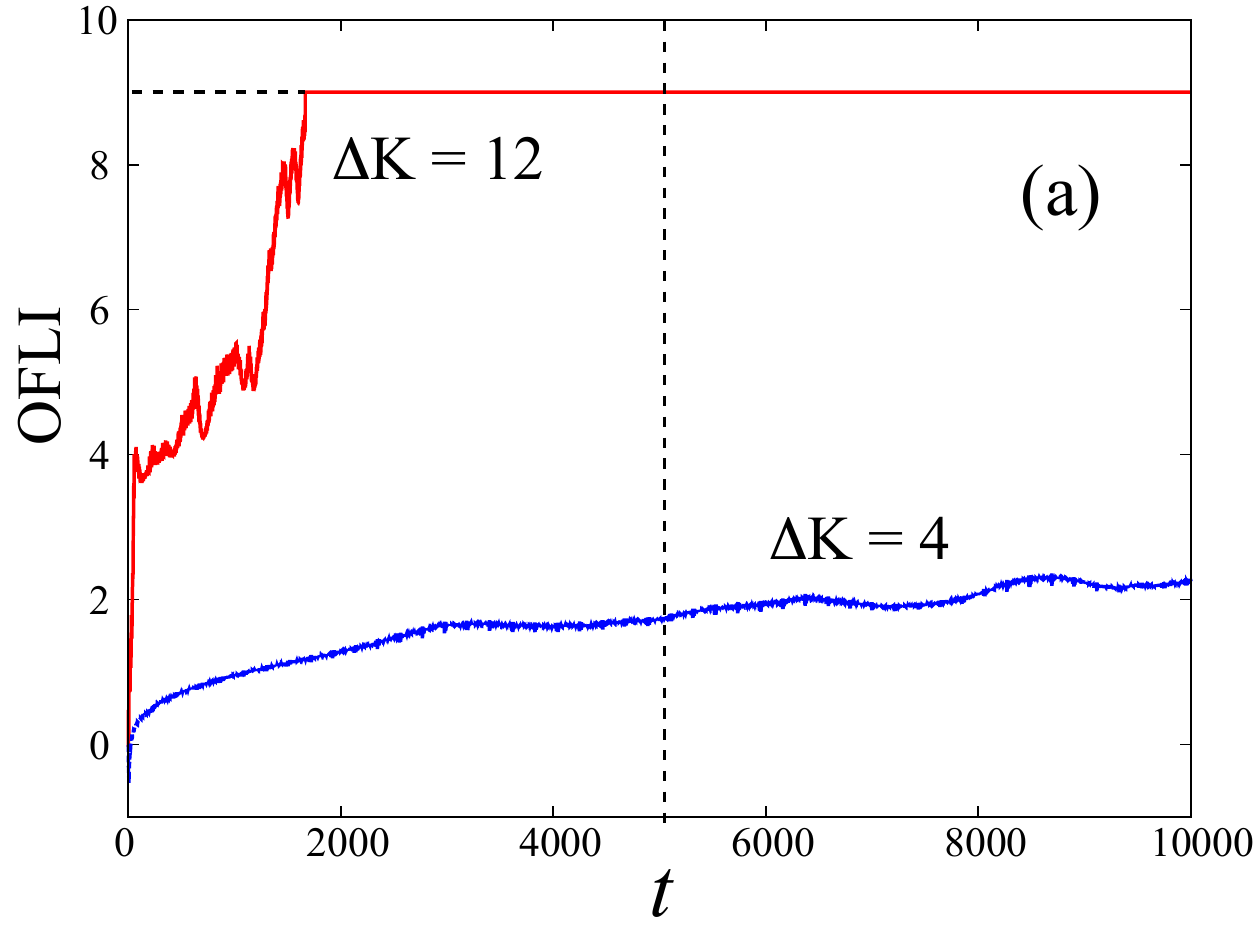}}

\centerline{\includegraphics[scale=0.8]{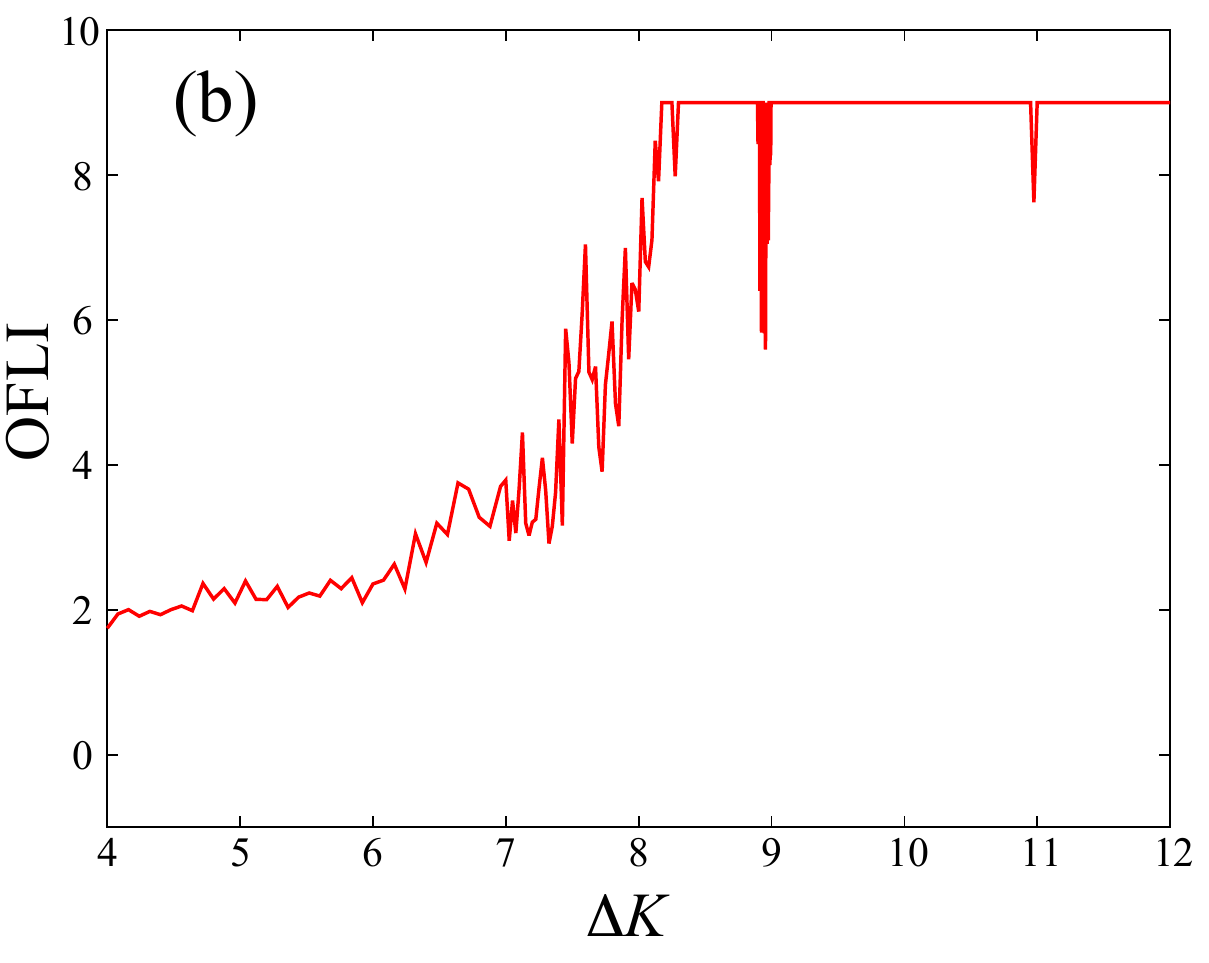}}

\caption{(a) Time evolution of the OFLI for $\Delta K$=4 (blue line) and $\Delta K$=12 (red line).
The vertical and horizontal black dashed lines indicate the cutoff value  of the OFLI computation and the
selected final time $t_f=5000$ in our simulations, respectively.
(b) The OFLI for a dipole chain with a single excited dipole as a function of the excitation energy $\Delta K$. Note that there is a cutoff value of the OFLI for the value 9.}
\label{fi:ofli1}
\end{figure}

As an example, we present in Fig.~\ref{fi:ofli1}(a)  the time evolution of the OFLI for two qualitatively different
orbits, belonging to the weakly (initial kinetic energy excess $\Delta K =4$) and to the highly 
(initial kinetic energy excess $\Delta K =12$) nonlinear regimes, respectively. 
We observe in Fig.~\ref{fi:ofli1}(a) that the OFLI for the $\Delta K =4$ trajectory increases very slowly, attaining
only small values (less than two up to very long times). In contrast the OFLI for the $\Delta K =12$ trajectory 
shows a fast increase reaching already at an early stage the 
cutoff value nine signifying its chaoticity. Note  that  at our usually selected final simulation time $t_f=5000$ the distinction between the two trajectories is clear, 
allowing for their classification as  regular ($\Delta K=4$) and chaotic ($\Delta K=12$)  respectively.

Our results for the behavior of the OFLI
 as a function of the excess energy $\Delta K$ are shown in Fig. \ref{fi:ofli1}(b).  In the regime
 $\Delta K \lesssim 6$, the value of the OFLI is below three, indicating the regular behavior of the system in accordance to Fig.~\ref{fi:meanC1} and  the above discussion. 
For
 $6 \lesssim \Delta K \lesssim 8$, the value
 of the OFLI increases for increasing energy, such that for $\Delta K \gtrsim 8.5$ it becomes larger
 than the cut-off value nine characterizing the chaotic orbits.
Moreover, the extracted chaoticity of
the dipole chain for excitations  with energy $\Delta K \gtrsim 8$ provides further
evidence for the link of the traveling energy localization patterns shown in Fig.~\ref{fi:colormap4} (c),(d)
and the striking phase locking states associated to them  (Fig.~\ref{fi:cosf1} (c),(d))
to chaotic breathers \cite{A851}, mentioned above.

As a final remark, it is worth noticing that, in general, the effectiveness of the fast Lyapunov indicators  in
providing a first indication of the degree of  chaoticity of an orbit with a relatively low
computational cost has been successfully proven in
dynamical systems with few degrees of freedom. Indeed, a global vision of the
phase space structure of several Hamiltonian systems with two or three degrees of freedom was obtained
by the computation of two-dimensional OFLI maps \cite{barrio, barrio1,PRE2014}. However, a detailed
investigation of the phase space and the chaotic dynamics of
multidimensional systems as our dipole chain requires the use of more sophisticated tools
based on the computation of Lyapunov exponents~\cite{osedelec,lichtemberg,skokos}
as deviation vector distributions~\cite{A908} and Lyapunov Weighted Dynamics~\cite{A911,A910}.
Although very interesting, these investigations go beyond the goal of the present study and
will be addressed in a future work.

\section{Conclusions}
We have explored the energy transfer mechanisms in a classical
dipole chain, modeled as  an array of $N$ rigid dipoles with their positions
fixed in space, interacting with their nearest neighbors and restricted to rotate in a common plane.
This leads to a
Hamiltonian system of $N$ degrees of freedom describing
the rotational dynamics of the dipole chain. The equilibrium points of the equations of motion
have been identified and analyzed. It turns out that these can be classified in several
families according to their stability, pointing to the high complexity of the potential energy surface of the chain of dipoles.
A linearization of the equations of motion around the GS configuration has lead to the harmonic approximation of the dipole chain Hamiltonian in terms of normal modes, a fact
that has allowed us to extract information about the linear spectrum of the  system.

The main focus of this work has been the study of the energy transfer of a localized 
excitation in the dipole chain for a varying excitation energy. Two regimes  with qualitatively different features have been identified. In the first regime of low energy excitations the system exhibits a weakly nonlinear behavior with the initially localized excitation spreading in the dipole chain, leading for large times to energy equipartition
among the dipoles.
The second regime of higher energy excitations is characterized by a strong nonlinearity 
causing energy localization in the form of discrete or chaotic breathers. In some cases the
formed chaotic breathers attain the character of domain walls, separating domains of dipoles 
with different polarization. This spontaneous phase locking of the dipole chain can be linked to the properties of the interaction potential and in particular to its lowest energy saddle point.

It turns out that in the highly nonlinear regime the dipole chain is very sensitive to the 
initial conditions, indicating its chaotic nature. To quantify the degree of chaoticity in the system for different values of the excitation energy we have calculated the Orthogonal
Fast Lyapunov Indicator which confirms the above discussed picture. For excitation energies below a certain
approximate threshold the dynamics is regular whereas
above it is highly nonlinear and chaotic.
Interestingly enough this threshold energy has a value close 
to the energy difference between the first saddle and the minimum of the interaction potential.

Suitable experimental realizations of our model could be provided 
by polar diatomic molecules trapped in a 1D optical lattice \cite{A753}, colloidal polar particles in optical
tweezers \cite{colloid_ref} or by rotating polar molecules in Helium nanodroplets \cite{helium_nano}.
Further theoretical studies
could be devoted to the investigation of the effect of an external homogeneous or inhomogeneous electric field on the dynamics and the energy transfer of a dipole chain as the one studied here.
Finally the exploration of the dynamics of the dipole chain in the full $2N$-dimensional case
would also be of interest owing to its even more complex potential landscape.

\appendix
\section{The character of the critical points}
The nature of the critical points can be judged by the eigenvalues of their Hessian matrix. Due to the
nearest-neighbor interactions  considered in this study, the Hessian
is almost tridiagonal, with an exception regarding the last element of the first row and first element of the last row  which are
different from zero due to the imposed PBC. 
In the following we discuss the character of the six families of  critical points of the dipole chain based on their Hessian eigenvalues.
\subsection{The minimum $E_m$ and the maximum $E_M$}
The $N\times N$ Hessian matrix of the critical point $\{x_i=0, \forall i\}$ or $\{x_i=\pi, \forall i\}$  takes the form:
\begin{equation} H_{min}=  \left( \begin{array}{ccccccc}
\label{hessian1}
4 &1 & 0 & 0 & ... & 0 & 1 \\
1 &4 & 1 & 0 & ... & 0 & 0 \\
0 &1 & 4 & 1 & ... & 0 & 0 \\
... &... & ... & ... & ... & ... & ... \\
0 &0 & 0 & 0 & ... & 4 & 1 \\
1 &0 & 0 & 0 & ... & 1 & 4
\end{array} \right).
\end{equation}
If $N$ is large, the last element of the first row and the first element of the last row have a negligible contribution to the eigenspectrum
of $H_{min}$. Thus, in terms of its eigenspectrum the Hessian can be approximated by the tridiagonal matrix 
\begin{equation} H_{min}\approx  \left( \begin{array}{ccccccc}
\label{hessian2}
4 &1 & 0 & 0 & ... & 0 & 0 \\
1 &4 & 1 & 0 & ... & 0 & 0 \\
0 &1 & 4 & 1 & ... & 0 & 0 \\
... &... & ... & ... & ... & ... & ... \\
0 &0 & 0 & 0 & ... & 4 & 1 \\
0 &0 & 0 & 0 & ... & 1 & 4
\end{array} \right).
\end{equation}
The $N$ eigenvalues of the matrices \ref{hessian1} and \ref{hessian2} are real
because $H_{min}$ is symmetric.
Moreover, the eigenvalues $\lambda_k$ of the tridiagonal matrix \ref{hessian2}
are given by~\cite{A862}
\[
\lambda_k=4 + 2 \cos\frac{k \ \pi}{(N+1)}.
\]
These eigenvalues are positive indicating  that also the eigenvalues of the exact matrix \ref{hessian1} would be positive 
and in turn that the corresponding critical point is a minimum. 

Returning to the original Hessian matrix \ref{hessian1}, the
analytic computation of its characteristic equation gives the following polynomial of degree $N$
\[
{\cal C}_{even} \equiv \lambda^N - a_{N-1} \lambda^{N-1} + a_{N-2} \lambda^{N-2}- ...-a_1 \lambda + a_0 =0,
\quad N\equiv \mbox{even},
\]
\[
{\cal C}_{odd} \equiv \lambda^N -a_{N-1} \lambda^{N-1} + a_{N-2} \lambda^{N-2}- ...+a_1 \lambda - a_0 =0, 
\quad N\equiv \mbox{odd},
\]
where $a_i>0$. This means that there are $N$ sign changes in
the sequence of the coefficients $(1, a_{N-1}, a_{N-2},..., a_1, a_0)$.
The rule of Descartes~\cite{descartes}, says that if $p$ is the number of
positive roots of a given polynomial and $s$ is the number of sign changes in the
coefficient sequence of this polynomial, then $s=p+2 k$, with
$k$ a positive integer. By virtue of this theorem, from the $N$
changes of sign in the coefficients $a_i$, we conclude that the Hessian~\ref{hessian1} has at
most $N$ positive eigenvalues. We can apply the Descartes rule also to extract information about the
maximum number of negative
roots. Indeed, after replacing $\lambda \rightarrow -\lambda$, the coefficients of the
odd degree monomials in $\lambda$ in the characteristic polynomials
${\cal C}_{even,odd}$ become negative, resulting in
 no sign changes in the coefficient sequence. This implies that there are no negative eigenvalues
and as a consequence all the $N$ eigenvalues of the exact Hessian~\ref{hessian1} are positive.

\medskip\noindent
For the critical point corresponding to  the
alternating configuration $\{x_i=\pi \left[1 \pm(-1)^i\right]/2, \forall i\}$, the Hessian matrix  takes the form:
\[ H_{max} = - H_{min}.
\]
with approximate eigenvalues
\[
\lambda_k=-(4 + 2 \cos\frac{k \ \pi}{(N+1)}),
\]
which are all negative,indicating that this critical point is a maximum. The
analytic computation of the characteristic equation yields
\[
{\cal C}_{max}=\lambda^N + a_{N-1} \lambda^{N-1} + a_{N-2} \lambda^{N-2}+ ...+a_1 \lambda + a_0 =0
\]
where $a_i<0$. Therefore, there are no sign changes in
the sequence of the coefficients $(1, a_{N-1}, a_{N-2},..., a_1, a_0)$. In this case, the Descartes rule
of signs assures that there are no positive roots of the characteristic polynomial.
If we replace $\lambda \rightarrow -\lambda$ in ${\cal C}_{max}$, the coefficients of the
odd degree monomials in $\lambda$ become negative, 
a fact that results in $N$ sign changes in the coefficient sequence. Thus the $N$ eigenvalues  of $H_{max}$ are negative
and the corresponding critical point is a maximum.

\subsection{The saddle points $E_{s1}$}
For the configurations made of  $b$ blocks of $n_j$ dipoles with
$x_k=\pi$ while the remaining  dipoles possess $x_i=0$, up to our 
knowledge, there exists no close expression for the  eigenvalues of the corresponding
almost tridiagonal Hessian matrix. However, from the numerical computation
of the characteristic equation for different values of $N$ and for
different number of blocks $b$, we have strong indications that these critical points
are saddle points of rank $2 b$ since, by applying the rule of Descartes, we find that the number of positive
and negative eigenvalues are $N-2 b$ and $2 b$, respectively.

\subsection{The saddle points $E_{s2}$}
For the two configurations  $\{x_i=\pi/2, \forall i\}$ or $\{x_i=-\pi/2, \forall i\}$, the Hessian matrix takes the form
\begin{equation} H_{s3}=  \left( \begin{array}{ccccccc}
\label{hessian3}
-2 &-2 & 0 & 0 & ... & 0 & -2 \\
-2 &-2 & -2 & 0 & ... & 0 & 0 \\
0 &-2 & -2 & -2 & ... & 0 & 0 \\
... &... & ... & ... & ... & ... & ... \\
0 &0 & 0 & 0 & ... & -2 & -2 \\
-2 &0 & 0 & 0 & ... & -2 & -2
\end{array} \right).
\end{equation}
Following~\cite{A862}, the approximate eigenvalues for large $N$ are given by the expression
\begin{equation}
\label{eigen3}
\lambda_k=-2 + 4 \cos\frac{k \ \pi}{(N+1)}.
\end{equation}
From Eq.~\ref{eigen3}, we obtain that this critical point is a saddle point with $2(N+1)/3$ hyperbolic
directions.

\subsection{The saddle points $E_{s3}$}
For the configuration with alternating angles $\pi/2$ and $-\pi/2$, $\{x_i= \pm(-1)^i\pi/2, \forall i\}$, the Hessian matrix takes the form
\begin{equation} H_{s3}=  \left( \begin{array}{ccccccc}
\label{hessian4}
2 &2 & 0 & 0 & ... & 0 & 2 \\
2 &2 & 2 & 0 & ... & 0 & 0 \\
0 &2 & 2 & 2 & ... & 0 & 0 \\
... &... & ... & ... & ... & ... & ... \\
0 &0 & 0 & 0 & ... & 2 & 2 \\
2 &0 & 0 & 0 & ... & 2 & 2
\end{array} \right).
\end{equation}
Following~\cite{A862}, the approximate eigenvalues for large $N$ are given by the expression
\begin{equation}
\label{eigen4}
\lambda_k=2 + 4 \cos\frac{k \ \pi}{(N+1)}.
\end{equation}
From Eq.~\ref{eigen4}, we see that this critical point is a saddle point with $(N+1)/3$ hyperbolic
directions.

\subsection{The critical points $E_{s4}$}
For the critical points made of all the possible configurations with $x_i=\pm \pi/2$, up to our 
knowledge, there  exists no close expression for the approximate eigenvalues of the corresponding
almost tridiagonal Hessian matrix. 
Moreover, from the numerical computation
of the characteristic equation for different configurations, we cannot conclude anything about
the nature of these critical points, because, depending on the equilibrium configuration, some
of the eigenvalues are zero.

\end{document}